\documentclass[aps,prb,superscriptaddress,twocolumn,showpacs]{revtex4-2}
\usepackage{amsmath}
\usepackage{amssymb}
\usepackage[pdftex]{graphicx}
\usepackage{bm}
\usepackage[pdftex]{color}
\usepackage{cases}
\usepackage{multirow}
\usepackage{braket}
\usepackage{mathtools}
\usepackage{comment}

\usepackage[final]{changes}

\def\stacksymbols #1#2#3#4{\def\theguybelow{#2}
    \def\verticalposition{\lower#3pt}
    \def\spacingwithinsymbol{\baselineskip0pt\lineskip#4pt}
    \mathrel{\mathpalette\intermediary#1}}
\def\intermediary#1#2{\verticalposition\vbox{\spacingwithinsymbol
      \everycr={}\tabskip0pt
      \halign{$\mathsurround0pt#1\hfil##\hfil$\crcr#2\crcr
               \theguybelow\crcr}}}

\usepackage[pdftex]{hyperref}
\begin{document}
\title{Dual Symmetry Classification of Non-Hermitian Systems and \\ $\mathbb{Z}_2$ Point-Gap Topology of a Non-Unitary Quantum Walk}

\author{Zhiyu Jiang}
\affiliation{Department of Applied Physics, Hokkaido University, Kita 13, Nishi 8, Kita-Ku, Sapporo, Hokkaido 060-8628, Japan}
\author{Ryo Okamoto}
\affiliation{Department of Electronic Science and Engineering, Kyoto University, Kyotodaigakukatsura, Nishikyo-ku, Kyoto 615-8510, Japan}
\author{Hideaki Obuse}
\affiliation{Department of Applied Physics, Hokkaido University, Kita 13, Nishi 8, Kita-Ku, Sapporo, Hokkaido 060-8628, Japan}
\affiliation{Institute of Industrial Science, The University of Tokyo, 5-1-5 Kashiwanoha, Kashiwa, Chiba 277-8574, Japan}


 \begin{abstract}
Non-Hermitian systems exhibit richer topological properties compared to their Hermitian counterparts. It is well known that non-Hermitian systems have been classified based on either the symmetry relations for non-Hermitian Hamiltonians or the symmetry relations for non-unitary time-evolution operators in the context of Floquet topological phases. In this work, we propose that non-Hermitian systems can always be classified in two ways; a non-Hermitian system can be classified using the symmetry relations for non-Hermitian Hamiltonians or time-evolution operator regardless of the Floquet topological phases or not. We refer to this as {\it dual symmetry classification}. To demonstrate this, we successfully introduce a new non-unitary quantum walk that exhibits point gaps with a $\mathbb{Z}_2$ point-gap topological phase applying the dual symmetry classification and treating the time-evolution operator of this quantum walk as the non-Hermitian Hamiltonian.
 \end{abstract}

\maketitle
\section{introduction}
\label{sec:introduction}

The study of non-Hermitian systems has gained significant attention in recent years due to its potential for novel physical phenomena, especially relating with topological phase, and practical applications\cite{Ashida-Review,Okuma-Review}. The eigenenergies of non-Hermitian Hamiltonians generically become complex. For classical systems, non-Hermitian Hamiltonians are frequently employed to describe optical systems with gain and loss\cite{PhysRevLett.106.213901,regensburger2012parity,feng2013experimental,weimann2017topologically,zhou2018observation,doi:10.1126/science.aay1064}, topological circuit systems\cite{PhysRevX.5.021031,PhysRevB.99.201411,PhysRevB.100.045407,helbig2020generalized,PhysRevResearch.2.022062,PhysRevResearch.2.023265,liu2021non}, etc. In open quantum systems, a non-Hermitian Hamiltonian can be used to describe the dynamics of a system without quantum jump process\cite{diehl2011topology,PhysRevLett.123.170401,PhysRevLett.124.040401,PhysRevA.105.053718}. In such cases, the system follows the Schrödinger equation with a non-Hermitian Hamiltonian.
 
In Ref.\cite{PhysRevX.9.041015}, the symmetry in non-Hermitian systems, which is important to the topological phases, has been studied by extending the symmetry relations for Hermitian Hamiltonians.
Then, they showed that there are 38 symmetry classes in non-Hermitian systems.
Furthermore,
In non-Hermitian systems, the energy gaps can be classified into two types: line gaps and point gaps\cite{PhysRevX.8.031079,PhysRevX.9.041015}. 
On one hand, the topological invariant for the line gap is related to the number of edge states appearing in the gap through the bulk-boundary correspondence (BBC) in the same fashion in Hermitian systems\cite{PhysRevLett.121.026808,PhysRevB.99.081103,PhysRevB.99.081302,PhysRevB.100.165430,PhysRevA.99.052118,xiao2020non,PhysRevResearch.2.013280,PhysRevLett.125.226402,PhysRevLett.126.216407,PhysRevB.103.075126,PhysRevB.108.L121302}. 

On the other hand, the point gap is unique to non-Hermitian systems and $\mathbb{Z}$ topological invariant defined as the winding number in the complex energy plane for the system with periodic boundary conditions (PBC) predicts that spectra for the system with open boundary conditions (OBC) collapse into arcs within the closed curves of the  PBC spectra. The corresponding eigenstates are localized near the open boundaries, known as the non-Hermitian skin effects(NHSE)\cite{PhysRevLett.123.170401,PhysRevB.100.054301,PhysRevLett.124.056802,Okuma2020,PhysRevLett.125.126402,li2020critical,PhysRevB.102.201103,li2020critical,PhysRevResearch.2.013280,PhysRevB.102.241202,zhang2021observation,PhysRevB.104.165117,PhysRevB.104.125109,PhysRevB.103.L140201,PhysRevB.103.085428,zhang2022universal,PhysRevB.106.235411,roccati2024hermitian,Nakai2024}. One of the famous examples is the Hatano-Nelson model\cite{PhysRevLett.77.570,PhysRevB.56.8651,PhysRevB.58.8384}.
It is also known that a system with spinful time-reversal symmetry exhibits $\mathbb{Z}_2$ topological invariant for the point gap\cite{Okuma2020}.

One promising area of exploration within topology of non-Hermitian physics is the study of non-unitary quantum walks\cite{PhysRevA.93.062116,xiao2017observation,kawasaki2020bulk,Mochizuki2020,Xiao2020,active-quantum-particle}. The discrete-time quantum walk consists of a walker moving on a lattice, obeying a unitary time-evolution operator with each time step\cite{PhysRevA.48.1687,tregenna2003controlling,PhysRevA.77.032326,PhysRevA.81.042330,PhysRevLett.104.050502,PhysRevB.88.121406,PhysRevA.107.042206}. However, the non-unitary quantum walks incorporate gain and loss, resulting in non-Hermitian Hamiltonian dynamics. It is known that the non-unitary time-evolution operator can be classified by translating the symmetry relations of Hamiltonians to time-evolution operators. Some models of non-unitary quantum walks have been considered theoretically\cite{PhysRevA.93.062116,kawasaki2020bulk,Mochizuki2020,active-quantum-particle} and realized experimentally\cite{xiao2017observation,Xiao2020} by temporally alternating photon losses. 
However, there is no research focused on the topological properties of time-reversal symmetric no-Hermitian system with non-trivial $\mathbb{Z}_2$ point-gap topology.

In this work, we introduce a non-unitary quantum walk with spinful time-reversal symmetry and $\mathbb{Z}_2$ point gap topology by proposing the alternative symmetry classification way which we call as dual symmetry classification. 
The new scheme of the classification is based on the fact that there is no strict difference in mathematical definition between non-Hermitian Hamiltonians and non-unitary time-evolution operators. This means that regardless of whether the system is defined by the Hamiltonian or the time-evolution operator, the non-Hermitian systems can be always classified by two ways; from the symmetry relations for non-Hermitian Hamiltonians and from those for non-unitary time-evolution operators.
By using this new classification scheme, we introduce a non-unitary quantum walk with spinful time-reversal symmetry.
We find that the non-unitary quantum walk has $\mathbb{Z}_2$ point gap with non-trivial $\mathbb{Z}_2$ topological invariant. We confirm that the non-unitary quantum walk satisfies BBC and the eigenstates are localized at both boundaries, indicating the $\mathbb{Z}_2$ NHSE.

This paper is organized as follows. In Sec.\ref{sec:symmetry}, we introduce our new way to define symmetry of time-evolution operators in non-unitary quantum walks. In Sec.\ref{sec:model}, we define a non-unitary quantum walk with spinful time-reversal symmetry$^{\dagger}$ and discuss the symmetry class of the system. Moreover, we propose methods of experimental realization of our quantum walk. In Sec.\ref{subsec:PBC and PBC} we calculate the $\mathbb{Z}_2$ point gap topological invariant of the non-unitary quantum walk and validate that when the quantum walk possesses a non-trivial $\mathbb{Z}_2$ topology, the eigenvalue spectrum of the time evolution operator under reflecting boundary conditions resides within the spectrum under periodic boundary conditions. Simultaneously, the eigenstates are localized near the boundaries of the system, indicating the emergence of the $\mathbb{Z}_2$ NHSE. In Sec.\ref{subsec:junction system}, we study a junction system comprised of two topologically non-trivial quantum walks with inhomogeneous parameters in real space. Intriguingly, we find that both subsystems exhibit skin effects even under periodic boundary conditions.

\section{Dual symmetry classification of non-Hermitian systems}
\label{sec:symmetry}

In Hermitian systems, a system is classified into the 10 AZ symmetry class by considering the presence or absence of three symmetries; time-reversal symmetry (TRS), partical-hole symmetry (PHS) and chiral symmetry (CS) or sublattice symmetry (SLS). In Ref.\cite{PhysRevX.9.041015}, because of the absence of Hermiticity $(H^{*}\neq H^{T})$, it is shown that the non-Hermitian systems possess the additional symmetry class called $\text{AZ}^{\dagger}$ symmetry class where TRS and PHS are ramified as TRS$^\dagger$ and PHS$^\dagger$, respectively. These symmetry relations are summarized as
\begin{subequations}
\label{eq:H}
\begin{align}
\mathcal{T}_+ H^{*}(k)\mathcal {T}_+ ^{-1} &=+H(-k)\quad\quad(\text{TRS}), \label{eq;TRS Hamiltonian} \\
\mathcal{T}_- H^{T}(k)\mathcal {T}_- ^{-1} &=+H(-k)\quad\quad(\text{TRS$^\dagger$}), \label{eq;TRS dagger Hamiltonian} \\
\Xi_+ H^{T}(k)\Xi_+^{-1} &=-H(-k)\quad\quad(\text{PHS}), \label{eq;PHS Hamiltonian} \\
\Xi_- H^{*}(k)\Xi_-^{-1} &=-H(-k)\quad\quad(\text{PHS$^\dagger$}), \label{eq;PHS dagger Hamiltonian} \\
\Gamma H^{\dagger}(k)\Gamma^{-1} &=-H(+k)\quad\quad(\text{CS}), \label{eq;CS Hamiltonian}\\
S H(k)S^{-1} &=-H(+k)\quad\quad(\text{SLS}), \label{eq;SLS Hamiltonian}
\end{align}
\end{subequations}
where $H(k)$ stands a non-Hermitian Hamiltonian in the momentum space and $\mathcal{T}_\pm, \Xi_\pm, \Gamma$, and $S$ are unitary operators.

In the study of topological phases, it is found that the non-trivial topological phase can be induced by periodically driving a system, which is known as the Floqeut topological phase. For the Floquet topological phase, symmetry should be defined through the time-evolution operator, not the instanton Hamiltonian since the time dependence of the Hamiltonian is crucial to retain the symmetry. 
Therefore, taking into account the relation between the Floquet effective Hamiltonian $H_{\text{eff}}(k)$ and the time-evolution operator $U(k)$, 
\begin{align}
U(k)&=e^{-iH_{\text{eff}}(k)},
\label{eq:H_eff}
\end{align}
the aforementioned relationships of symmetry can be reformulated as
\begin{subequations}
\label{eq:U}
\begin{align}
\mathcal{T}_+ U^{*}(k)\mathcal {T}_+ ^{-1} &=U^{-1}(-k)\quad\quad(\text{TRS}), \label{eq;TRS time evolution operator} \\
\mathcal{T}_- U^{T}(k)\mathcal {T}_-^{-1} &=U(-k),\quad\quad\ \ (\text{TRS$^\dagger$}) \label{eq;TRS dagger time evolution operator} \\
\Xi_+ U^{T}(k)\Xi_+^{-1} &=U^{-1}(-k)\quad\quad(\text{PHS}), \label{eq;PHS time evolution operator} \\
\Xi_- U^{*}(k)\Xi_-^{-1} &=U(-k)\quad\quad\quad(\text{PHS$^\dagger$}), \label{eq;PHS dagger time evolution operator} \\
\Gamma U^{\dagger}(k)\Gamma^{-1} &=U(k)\quad\quad\quad\ \ (\text{CS}), \label{eq;CS time evolution operator}\\
S U(k)S^{-1} &=U^{-1}(k)\quad\quad\ \ (\text{SLS}). \label{eq;SLS time evolution operator}
\end{align}
\end{subequations}
where $U(k)$ stands for a non-unitary time-evolution operator in the momentum space.
The quasi-energy is defined by the eigenvalue equation of $U(k)$
\begin{equation}
U|\phi\rangle=\lambda|\phi\rangle =e^{-i \varepsilon}|\phi\rangle.
\end{equation}
For a non-unitary time-evolution operator, in general, $|\lambda|\ne 1$ and $\varepsilon$ is complex. 

While it is well known that the classifications based on the above symmetry relations in Eqs.\ (\ref{eq:H}) and (\ref{eq:U}) correctly predicts non-trivial topological phases in non-Hermitian systems, 
here, we propose alternative way to classify the non-Hermitian systems based on the fact that there is no strict difference in mathematical definition between non-Hermitian Hamiltonians and non-unitary time-evolution operators. By regarding a given non-Hermitian Hamiltonian $H(k)$ as a non-unitary time-evolution operator $U(k)$, the non-Hermitian Hamiltonian can be classified by symmetries defined in Eq.\ (\ref{eq:U}). In the same way, regarding a non-unitary time-evolution operator $U(k)$ as an non-Hermitian Hamiltonian $H(k)$, the non-unitary time-evolution operator can be classified by Eq.\ (\ref{eq:H}). 
This means that there are always two ways to classify a given Hamiltonian or a time-evolution operator in non-Hermitian systems, which we call dual symmetry classification.
For example, the symmetry class of a give time-evolution operator is determined by the relations in Eqs.\ (\ref{eq;TRS Hamiltonian})-(\ref{eq;SLS Hamiltonian}) or the relations in Eqs.\ (\ref{eq;TRS time evolution operator}) - (\ref{eq;SLS time evolution operator}).
To avoid confusion, we call the former (latter) scheme for the classification as the $H$-type ($U$-type) classification. For the sake of completeness, we summarize all the symmetry relations in Table \ref{tab:symmetry}.

Note that Eq.\ (\ref{eq;TRS dagger Hamiltonian}) and Eq.\ (\ref{eq;TRS time evolution operator}) for TRS$^\dagger$ are the same. In addition, Eq.\ (\ref{eq;TRS Hamiltonian}) for TRS and Eq.\ (\ref{eq;PHS dagger time evolution operator}) for PHS$^\dagger$ are the same.
We also note the inverse of the operator on the right-hand side in Eqs.\ (\ref{eq;TRS time evolution operator}), (\ref{eq;PHS time evolution operator}), and (\ref{eq;SLS time evolution operator}) for $U$-type symmetry requires the strong constraints on the Hamiltonian. However, the minus sign on the right-hand side in Eqs.\ (\ref{eq;PHS Hamiltonian})-(\ref{eq;SLS Hamiltonian}) for $H$-type symmetry requires the time-evolution operator only shifting the quasi-energy by $\pi$.

In the subsequent sections, we demonstrate the validity of the dual symmetry classification by considering a non-unitary time-evolution operator defined for a quantum walk.

\begin{table*}[t]
\caption{A list of the symmetry relations. For the dual symmetry classification we represent either a non-Hermitian Hamiltonian or a non-unitary time-evolution operator for the non-Hermitian system as $X(k)$.}
\label{tab:symmetry}
\begin{tabular}[t]{l l l l l }
\hline
 & Ref.\ \cite{PhysRevX.9.041015} &Ref.\ \cite{kawasaki2020bulk} & dual symmetry classification &   \\
Symmetry$\quad$  & Hamiltonian & Time-evolution op. & $H$ type & $U$ type \\
\hline
TRS & $\mathcal{T}_+ H^{*}(k)\mathcal {T}_+ ^{-1} =H(-k)\quad\quad$ & $\mathcal{T}_+ U^{*}(k)\mathcal {T}_+ ^{-1} =U^{-1}(-k)\quad\quad$&$\mathcal{T}_+ X^{*}(k)\mathcal {T}_+ ^{-1} =X(-k)\quad\quad$ & $\mathcal{T}_+ X^{*}(k)\mathcal {T}_+ ^{-1} =X^{-1}(-k)$\\
TRS$^\dagger$ & $\mathcal{T}_- H^{T}(k)\mathcal {T}_- ^{-1} =H(-k)$ & $\mathcal{T}_- U^{T}(k)\mathcal {T}_-^{-1} =U(-k)$ &$\mathcal{T}_- X^{T}(k)\mathcal {T}_- ^{-1} =X(-k)\quad\quad$& $\mathcal{T}_- X^{T}(k)\mathcal {T}_-^{-1} =X(-k)$ \\
PHS & $\Xi_+ H^{T}(k)\Xi_+^{-1} =-H(-k)$ & $\Xi_+ U^{T}(k)\Xi_+^{-1} =U^{-1}(-k)$& $\Xi_+ X^{T}(k)\Xi_+^{-1} =-X(-k)\quad\quad$ & $\Xi_+ X^{T}(k)\Xi_+^{-1} =X^{-1}(-k)$ \\
PHS$^\dagger$ & $\Xi_- H^{*}(k)\Xi_-^{-1} =-H(-k)$ & $\Xi_- U^{*}(k)\Xi_-^{-1} =U(-k)$ & $\Xi_- X^{*}(k)\Xi_-^{-1} =-X(-k)\quad\quad$ & $\Xi_- X^{*}(k)\Xi_-^{-1} =X(-k)$ \\
CS & $\Gamma H^{\dagger}(k)\Gamma^{-1} =-H(k)$ & $\Gamma U^{\dagger}(k)\Gamma^{-1} =U(k)$ & $\Gamma X^{\dagger}(k)\Gamma^{-1} =-X(k)\quad\quad$ & $\Gamma X^{\dagger}(k)\Gamma^{-1} =X(k)$ \\
SLS & $S H(k)S^{-1} =-H(k)$ & $S U(k)S^{-1} =U^{-1}(k)$ &$S X(k)S^{-1} =-X(k)\quad\quad$& $S X(k)S^{-1} =X^{-1}(k)$ \\
\hline
\end{tabular}
\end{table*}
\section{Model}
\label{sec:model}

In this section, we introduce a non-unitary discrete-time quantum walk to demonstrate the dual symmetry classification for the time-evolution operator.
Due to the reason explained in Sec.\ \ref{sec:point-gap topologyof the non-unitary quantum walk}, we call this quantum walk the non-unitary quantum walk with $\mathbb{Z}_2$ point-gap topology.

\subsection{Real space representation}
We consider a particle with two internal degrees of freedom, which is called a walker, moving on a pair of parallel chains, $a$ and $b$, which is important to realize TRS$^\dagger$ with $\mathcal{T}_{-}\mathcal{T}_{-}^*=-1$, as we demonstrate later.
The state of the walker can be defined as $\left|\psi(t)\right\rangle=\sum_{x,n,s}\psi_{x,n,s}(t)\left|x\right\rangle\otimes\left|n\right\rangle\otimes\left|s\right\rangle$. Here, $|x\rangle$ where $x\in \mathbb{Z}$ represents the position space, $|n=a,b\rangle$ represents the chain space, and $|s=L, R\rangle$ represents the internal state space where two basic vectors of the internal state of the walker is $\left| L\right\rangle=\left(1,0\right)^{T}$ and $\left| R\right\rangle=\left(0,1\right)^{T}$. The dynamics of a quantum walk are described by a time-evolution operator $U$ periodically operating on the wave function of the walker. After $t$ steps, the state of walker can be written as $|\psi(t)\rangle=U^{t}|\psi(0)\rangle$.

\begin{figure}[tb]
  \centering
  \includegraphics[width=0.8\columnwidth]{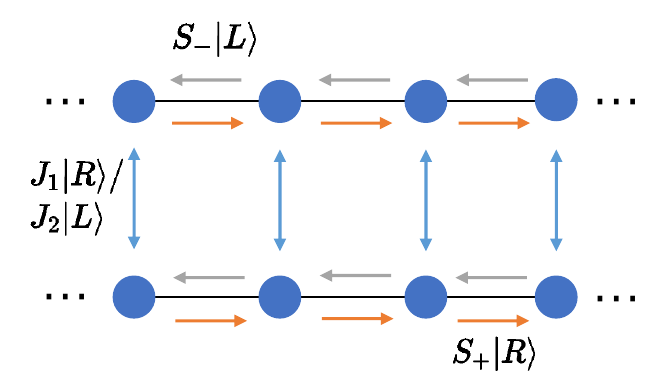}
  \caption{The schematic view of the movement of the walker in the non-unitary quantum walk in Eq.\ (\ref{eq:time evolution operator}). The walker is moving on a pair of parallel chains. The horizontal movement of the walker along the chain is controlled by $S$ operator, while the jump to the other chain is controlled by $J$ operator. }
  \label{fig:model}
\end{figure}

A standard quantum walk within one period involves at least two operators: the coin operator and the shift operator. These operators serve to simulate the processes of coin flipping and the movement of the walker, respectively.
We define the coin operator $C$ as a rotation operation in the opposite direction of the internal state on different chains, which can be expressed as
\begin{align}
C_{\alpha=x,y}[\theta(x)]&=\sum_{x}|x\rangle\langle x|\otimes e^{-i\theta(x)\tau_z\otimes\sigma_{\alpha}} \\
&=\sum_{x}|x\rangle\langle x|\otimes
\begin{pmatrix}
e^{-i\theta(x)\sigma_\alpha}&0 \\
 0&e^{i\theta(x)\sigma_\alpha}
\label{eq:coin operator}
\end{pmatrix},
\end{align}
where the Pauli matrix $\sigma_{x,y,z}$ ($\tau_{x,y,z}$) acts on the internal states space $|s\rangle$ (the chain space $|n\rangle$).
The subscript $\alpha = x,y$ indicates the rotation around the $x$ or $y$ axis with the rotation angle $\theta(x)$ which may depend on the position $x$.

The movement of the walker along the same chain is determined by the shift operators. The shift operators change the position $|x\rangle$ depending on the initinal states, are given by
\begin{align}
S_{-}&=\sum_{x}\left( |x-1\rangle\langle x|\otimes \tau_0\otimes|L\rangle\langle L|+|x\rangle\langle x|\otimes \tau_0\otimes|R\rangle\langle R| \right),\label{eq:shift operator -}
\\
S_{+}&=\sum_{x}\left( |x\rangle\langle x|\otimes \tau_0\otimes|L\rangle\langle L|+|x+1\rangle\langle x|\otimes \tau_0\otimes|R\rangle\langle R| \right),
\label{eq:shift operator}
\end{align}
where $\tau_0$ represent the $2\times 2$ indentity matrix acting on the chain space.
Here, we take periodic boundary condition to ensure that the shift operator is unitary.

We introduce another operator representing the hopping between the two chains that we refer to as the jump operator $J_{1,2}$. This operator determines whether the walker jumps to another chain depending on the internal state. The jump operators $J_{1,2}$ can be expressed as follows:
\begin{align}
J_{1}&=\sum_{x} |x\rangle\langle x|\otimes\left[\tau_0\otimes|L\rangle\langle L|+(|b\rangle\langle a|+|a\rangle\langle b|)\otimes|R\rangle\langle R| \right],\\
J_{2}&=\sum_{x} |x\rangle\langle x|\otimes\left[ (|b\rangle\langle a|+|a\rangle\langle b|)\otimes|L\rangle\langle L|+ \tau_0\otimes|R\rangle\langle R| \right].
\label{eq:jump operator}
\end{align}
The movement of the walker by the shift operator $S_\pm$ and the jump operator $J_{1,2}$ are schematically shown in Fig.\ref{fig:model}. 

To incorporate non-unitary dynamics into the quantum walk, we introduce a non-unitary gain-loss operator
\begin{align}
G&=\sum_{x}|x\rangle\langle x|\otimes e^{\gamma\tau_z\otimes\sigma_z}\\
&=\sum_{x}|x\rangle\langle x|\otimes
\begin{pmatrix}
e^{\gamma\sigma_z}&0 \\
0 &e^{-\gamma\sigma_z}
\end{pmatrix},
\label{eq:non-unitary operator}
\end{align}
which accounts for the phenomenological gain and loss of amplitudes, while $\gamma$ represents the strength of non-unitarity of the quantum walk.

Using the above operators, we define the time evolution operator of our quantum walk in a time-symmetric frame as
\begin{equation}
\begin{split}
U=&C_y\left[\theta_3(x)/2\right]S_{+}C_y[\theta_1(x)]J_2C_x[\theta_2(x)]G \\
&C_x[\theta_2(x)]J_1C_y[\theta_1(x)]S_{-}C_y\left[\theta_3(x)/2\right].
\label{eq:time evolution operator}
\end{split}
\end{equation}
When we set the value of $\theta_3$ at the boundary to be $\pi/2$, the walker cannot move from one side boundary to the other. In this case, the boundary conditions are referred to as reflecting boundary conditions (RBC), which can also be viewed as OBC in a tight-binding model.

\subsection{Momentaum space representation}

If the parameter $\theta(x)$ in Eq.\ (\ref{eq:coin operator}) and $\gamma$ in Eq.\ (\ref{eq:non-unitary operator}) are constant over the position $x$, the time-evolution operator $U$ possesses translation symmetry and can be diagonal in the momentum space $|k\rangle$ by the Fourier transformation of the basic operators over the real space $|x\rangle$, yielding:
\begin{equation}
C_{\alpha}[\theta]=\sum_{k}|k\rangle\langle k|\otimes \widetilde{C}_{\alpha}(\theta),\ \widetilde{C}_{\alpha}(\theta)=
\begin{pmatrix}
e^{-i\theta\sigma_{\alpha}}&0 \\
 0&e^{i\theta\sigma_{\alpha}}
\end{pmatrix},
\label{eq:coin operator k}
\end{equation}
\begin{align}
S_{-}&=\sum_{k}|k\rangle\langle k|\otimes \widetilde{S}_{-}(k),\ \widetilde{S}_{-}(k)=
\begin{pmatrix}
e^{ik}& & & \\
  &1 & & \\
 & &e^{ik}& \\
 & & &1\\
\end{pmatrix},\\
S_{+}&=\sum_{k}|k\rangle\langle k|\otimes \widetilde{S}_{+}(k),\ \widetilde{S}_{+}(k)=
\begin{pmatrix}
1& & & \\
  &e^{-ik} & & \\
 & &1& \\
 & & &e^{-ik}\\
\end{pmatrix},
\label{eq:shift operator k}
\end{align}
\begin{align}
J_{1}&=\sum_{k}|k\rangle\langle k|\otimes \widetilde{J}_{1},\ \widetilde{J}_{1}=
\begin{pmatrix}
0&0&1&0\\
0&1&0&0\\
1&0&0&0\\
0&0&0&1\\
\end{pmatrix},\\
J_{2}&=\sum_{k}|k\rangle\langle k|\otimes \widetilde{J}_{2},\ \widetilde{J}_{2}=
\begin{pmatrix}
1&0&0&0\\
 0&0&0&1\\
0&0&1&0\\
0&1&0&0\\
\end{pmatrix},
\label{eq:jumping operator k}
\end{align}
\begin{equation}
G=\sum_{k}|x\rangle\langle k|\otimes\ \widetilde{G},\ \widetilde{G}=
\begin{pmatrix}
e^{\gamma\sigma_z}&0 \\
0 &e^{-\gamma\sigma_z}
\end{pmatrix},
\label{eq:non-unitary operator k}
\end{equation}
with time evolution operator
\begin{equation}
U=\sum_{k}|k\rangle\langle k|\otimes \widetilde{U}(k), 
\end{equation}
\begin{equation}
\begin{split}
\widetilde{U}(k)=&\widetilde{C}_y\left(\theta_3/2\right)\widetilde{S}_{+}\widetilde{C}_y(\theta_1)\widetilde{J}_2\widetilde{C}_x(\theta_2)\widetilde{G} \\
& \widetilde{C}_x(\theta_2)\widetilde{J}_1\widetilde{C}_y(\theta_1)\widetilde{S}_{-}\widetilde{C}_y\left(\theta_y/2\right).
\label{eq:time evolution operator}
\end{split}
\end{equation}

\begin{figure}[tb]
  \centering
  \includegraphics[width=\columnwidth]{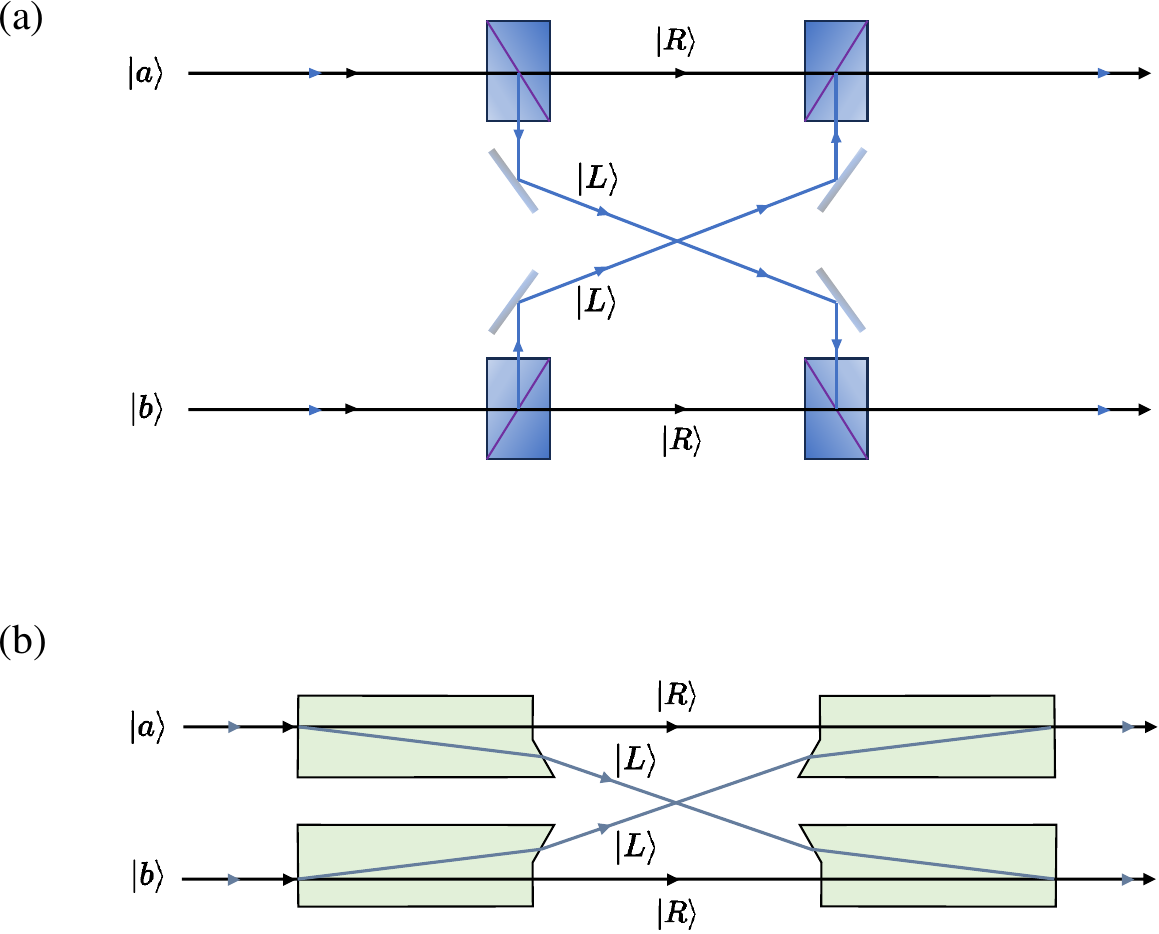}
  \caption{The schematic view of the two experimental implementation for the jump operator $J_2$ using (a) beam splitters and (b) beam displacers. The walker with an internal state of $|L\rangle$ undergoes a jump to the other chain, while the walker with an internal state of $|R\rangle$ remains on the same chain. $J_1$ can be implemented by rotating the polarization of the photons entering each optical path of the $J_2$ setup by 90 degrees using a half-wave plate and the polarization of the photons output from the $J_2$ setup by 90 degrees using a half-wave plate.}
  \label{fig:experiment}
\end{figure}

\subsection{Symmetry}
Here, we check the symmetry of the non-unitary time-evolution operator $\tilde{U}(k)$ in Eq.\ (\ref{eq:time evolution operator}) according to our proposal in Sec. \ref{sec:symmetry}.
First, we observe that the operator $\widetilde{U}(k)$ fulfills spinful TRS$^\dagger$ in $H$ type in Eq.\ \eqref{eq;TRS dagger Hamiltonian} with $\mathcal{T}_-=\tau_x\otimes \sigma_y$ where $\mathcal{T}_-\mathcal{T}_-^* = -1$.  Thus, the operator $\tilde{U}(k)$ also satisfies the TRS$^\dagger$ in $U$ type in Eq.\ (\ref{eq;TRS dagger time evolution operator}) since both relations are identical. Due to spinful TRS$^\dagger$, the eigenvalue of $\tilde{U}(k)$ is always double degenerated.

Furthermore, the operator $\tilde{U}(k)$ satisfies PHS$^\dagger$ in $H$ type, namely Eq.\eqref{eq;PHS dagger Hamiltonian} with $\Xi_-=\tau_z\otimes \sigma_0$.
Thereby, CS in $H$ type in Eq.\eqref{eq;CS Hamiltonian} with $\Gamma=\tau_y\otimes \sigma_y$ is also satisfied. 
Accordingly, the operator $\tilde{U}(k)$ is classified in class DIII$^\dagger$ in $H$ type, though  it is classified in class AII$^\dagger$ in $U$ type.

We remark that the symmetry realizations in Eq.\ (\ref{eq;PHS dagger Hamiltonian}) assumes that the eigenenergy $E$ appears a pair $\pm E$, thus the particle-hole symmetric point $E_0=0$.
If the symmetric point $E_0$ is not zero, Eq.\ (\ref{eq;PHS dagger Hamiltonian}) is modified as
\begin{equation}
\Xi_- \left[H(k) - E_0\right]^{*}\Xi_-^{-1} =-\left[H(-k) - E_0\right]. \label{eq;PHS dagger Hamiltonian mod}
\end{equation}
This means that when the operator satisfies Eq.\ (\ref{eq;PHS dagger Hamiltonian}), the operator also satisfies Eq.\ (\ref{eq;PHS dagger Hamiltonian mod}) only for $\text{Re}\left[E_0\right] =0$. In other words, PHS$^\dagger$ is broken for $\text{Im}\left[E_0\right] \ne 0$. Accordingly, our quantum walk $\tilde{U}(k)$ belongs in class AII$^\dagger$ even in $H$ type if the symmetric point is not a pure imaginary number. While this fact may imply that the dual symmetry classification is not important, this is not the case as we explained in Sec.\ \ref{sec:point-gap topologyof the non-unitary quantum walk}.

We also note that while we consider symmetry of our quantum walk only in the momentum representation for simplicity, symmetries mentioned above hold even in the time-evolution operator in the real space. Therefore, even if the parameter $\theta(x)$ in Eq.\ (\ref{eq:coin operator}) and $\gamma$ in Eq.\ (\ref{eq:non-unitary operator}) depend on the position, the symmetry class is unchanged.

\subsection{Experimental implementation}


At the end of Sec.\ \ref{sec:model}, we simply mention the experimental realization of our quantum walk.
In the experiment of photonic quantum walks in non-unitary dynamics, the coin operator, shift operator, and gain-loss operator can be realized by using a half-wave plate, beam displacer, and partially polarizing beam splitter, respectively\cite{xiao2017observation}. To experimentally implement the jump operator in our model, we propose two methods utilizing distinct optical setups, as illustrated in Fig. \ref{fig:experiment}. In Fig. \ref{fig:experiment}(a), we employ polarizing beam splitters and mirrors to facilitate the jump of photons possessing a specific polarization to the other chain. Here, photons with the $|R\rangle$ internal state traverse the polarizing beam splitter directly, maintaining their original direction. Conversely, photons with the $|L\rangle$ internal state undergo a $\pi/2$ reflection. After two subsequent reflections in the mirrors, they traverse the beam splitter of the alternate chain in a direction parallel to the incident path. In Fig. \ref{fig:experiment}(b), a notched rectangle beam displacer is employed. This architecture is expected to be more compact and stable than that in Fig. \ref{fig:experiment}(a). Photons with the $|R\rangle$ internal state remain on their original chain, while photons with the $|L\rangle$ internal state undergo displacement. After two refractions, they move to the alternate chain in a direction parallel to the original chain.

\section{$\mathbb{Z}_2$ point-gap topology of the non-unitary quantum walk}
\label{sec:point-gap topologyof the non-unitary quantum walk}

In this section, we study the point-gap topology of our non-unitary quantum walk $\tilde{U}(k)$ in Eq.\ \ref{eq:time evolution operator}.
We recall that the operator $\tilde{U}(k)$ belongs to class DIII$^\dagger$ in $H$ type for the symmetric point $\text{Re}[E_0]=0$, otherwise it belongs to class AII$^\dagger$ in $H$ type.
On the other hand, the operator $\tilde{U}(k)$ belongs to class AII$^\dagger$ in $U$ type for any symmetric point.

\subsection{$\mathbb{Z}_2$ point-gap topology under PBC and RBC}
\label{subsec:PBC and PBC}
\begin{figure}[tb]
  \centering
  \includegraphics[width=\columnwidth]{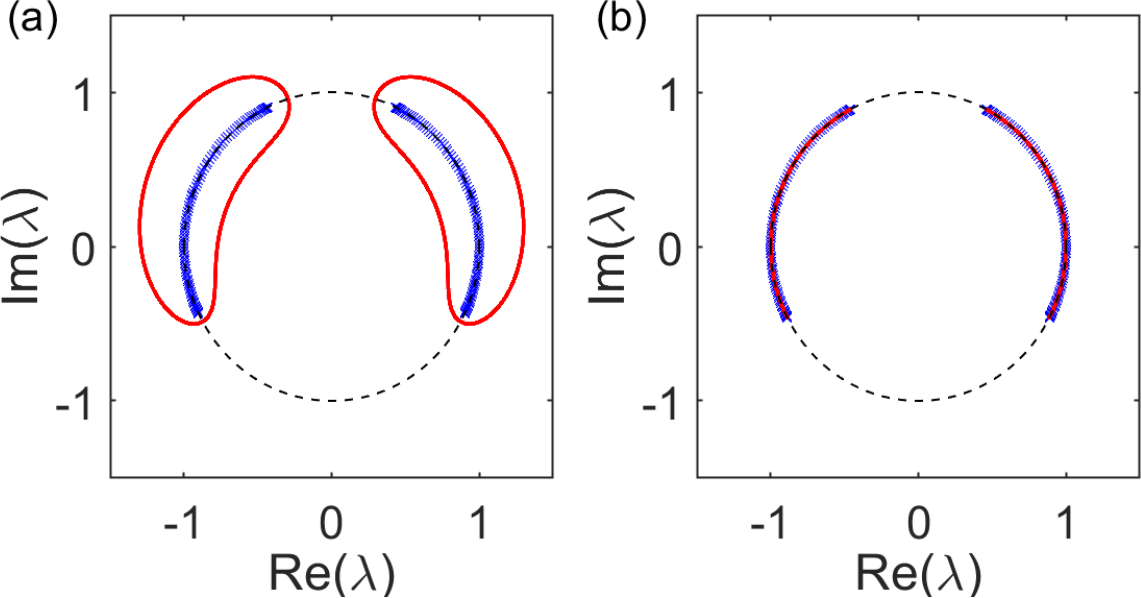}
  \caption{Eigenvalue spectra of the time evolution operator $U$ under PBC (red curves) and RBC (blue dots) in the case of $\gamma=0.5$, $\theta_1=3\pi/10$ and $\theta_3=\pi/4$. (a) $\theta_2=\pi/5$, (b) $\theta_2=\pi/6$. The black dashed line represents the unit circle. }
  \label{fig:spectra}
\end{figure}

The eigenvalues $\lambda$ of $\widetilde{U}(k)$ are shown by red curves in Fig.\ \ref{fig:spectra} for two specific values of parameters. Figure\ \ref{fig:spectra}(a) clearly shows that the point gaps open in finite regions of the symmetric point $E_0$, while the point gaps close by slightly changing the value of $\theta_2$ from $\pi/5$ to $\pi/6$ as shown in Figure\ \ref{fig:spectra}(b). 
However, due to  the double degeneracy of the eigenvalue, the energy-winding number for class DIII$^\dagger$ is always trivial. However, an additional topological invariant, known as $\mathbb{Z}_2$ topological invariant, defined in Ref. \cite{PhysRevX.9.041015} by
\begin{equation}
(-1)^{\nu(E_0)} =\text{sgn}\left\{\frac{\text{Pf}\left[\left(\widetilde{U}(\pi)-E_0\right) \mathcal{T}\right]}{\text{Pf}\left[\left(\widetilde{U}(0)-E_0\right) \mathcal{T}\right]}\right\},
\label{eq:Z_2 invariant}
\end{equation}
can be topologically non-trivial for $\text{Re}\left[E_0\right]=0$ in class DIII$^\dagger$ in $H$ type\replaced{, where}{.
Here,} $\nu\in\{0,1\}$ represents $\mathbb{Z}_2$ topological invariant, with $\nu=0$ indicating a topologically trivial phase.
\deleted{We confirmed that $\nu=1$ when $E_0$ locates in the closed curves in Fig.\ \ref{fig:spectra}(a).}
Here, we emphasize that we use the non-unitary time-evolution operator $\tilde{U}(k)$ instead of the effective Hamiltonian in Eq.\ (\ref{eq:Z_2 invariant}) to calculate the $\mathbb{Z}_2$ topological invariant while Eq.\ (\ref{eq:Z_2 invariant}) is defined for non-Hermitian Hamiltonians. Therefore, the fact that the regarding the time-evolution operator as the non-Hermitian Hamiltonian, which is crucial for the dual symmetry classification, is essential to study the topological phase.

However, as we mentioned, our quantum walk belongs to class AII$^\dagger$ in $H$ type for general value of $E_0$.
In this case, $\mathbb{Z}_2$ topological invariant at a symmetric point $E_0$ is defined by
\begin{equation}
\begin{aligned}
(-1)^{\nu(E_0)} & =\operatorname{sgn}\left\{\frac{\operatorname{Pf}\left[\left(\widetilde{U}(\pi)-E_0\right) \mathcal{T}\right]}{\operatorname{Pf}\left[\left(\widetilde{U}(0)-E_0\right) \mathcal{T}\right]}\right. \\
& \left.\times \exp \left[-\frac{1}{2} \int_{k=0}^{k=\pi} d k \frac{\partial \log \operatorname{det}\left[\left(\widetilde{U}(k)-E_0\right) \mathcal{T}\right]}{\partial k}\right]\right\}
\label{eq:Z2topo}
\end{aligned}
\end{equation}
in Ref. \cite{PhysRevX.9.041015}. \added{We confirmed that $\nu=1$ when $E_0$ locates in the closed curves in Fig.\ \ref{fig:spectra}(a).}
Again we remark that we use $\tilde{U}(k)$, not the Hamiltonian, to calculate the $\mathbb{Z}_2$ topological invariant.
Since Eq.\ (\ref{eq:Z2topo}) is defined for the operator in $H$ type, if $\mathbb{Z}_2$ invariant can be correctly calculated in Eq.\ (\ref{eq:Z2topo}), this clarifies the correctness of our proposal on the dual symmetry classification.

\begin{figure}[tb]
  \centering
  \includegraphics[width=\columnwidth]{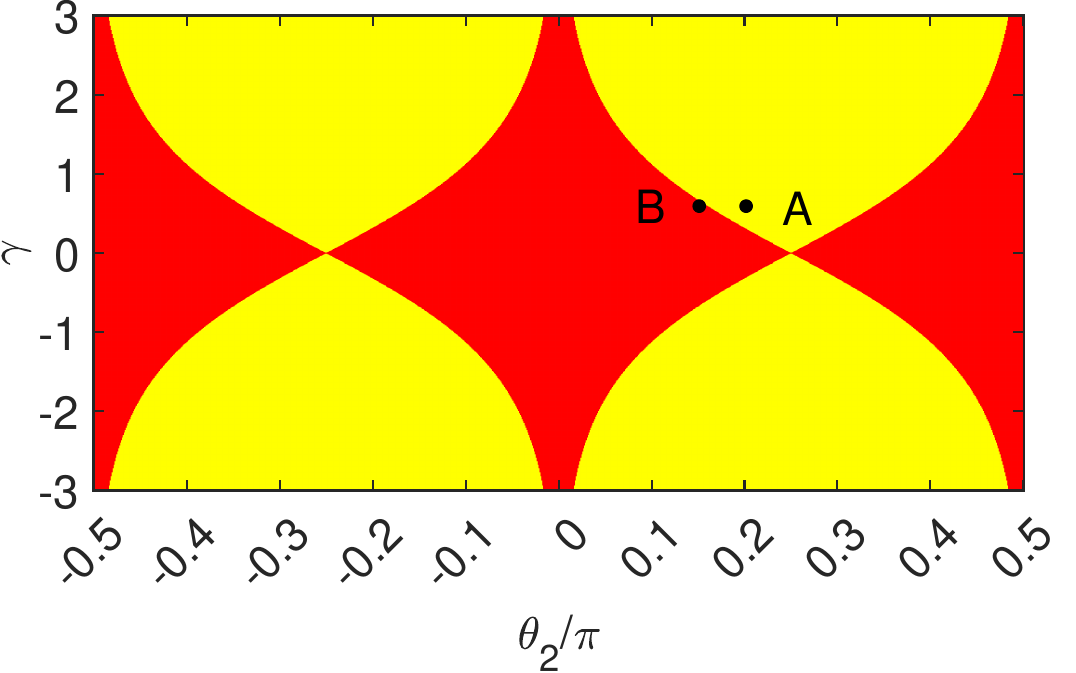}
  \caption{The phase diagram of $\mathbb{Z}_2$ topological invariant in the parameter space $(\theta_2,\gamma)$, with $\theta_1=\frac{3}{10}\pi$ and $\theta_3=\frac{\pi}{4}$. In the region depicted by filled yellow, $\nu(\lambda)=1$ for some values of $\lambda$, while in the region shown by filled red, $\nu(\lambda)\equiv0$ for any value of $\lambda$.}
  \label{fig:phase diagram}
\end{figure}

 Figure\ \ref{fig:phase diagram} illustrates the $\mathbb{Z}_2$ topological invariant of $\widetilde{U}(k)$ as a function of the parameters $\theta_2$ and $\gamma$ and the symmetric point $E_0$ by numerically solving Eq.(\ref{eq:Z2topo}). The yellow region represents the presence of at least one point $E_0\in\mathbb{C}$ where $\nu(\lambda)=1$, while the blue region indicates that any points on the complex plane is topologically trivial. The A and B in Fig.\ \ref{fig:phase diagram} indicates the parameters used in Fig.\ \ref{fig:spectra} (a) and (b), respectively.

\begin{figure}[tb]
  \centering
  \includegraphics[width=\columnwidth]{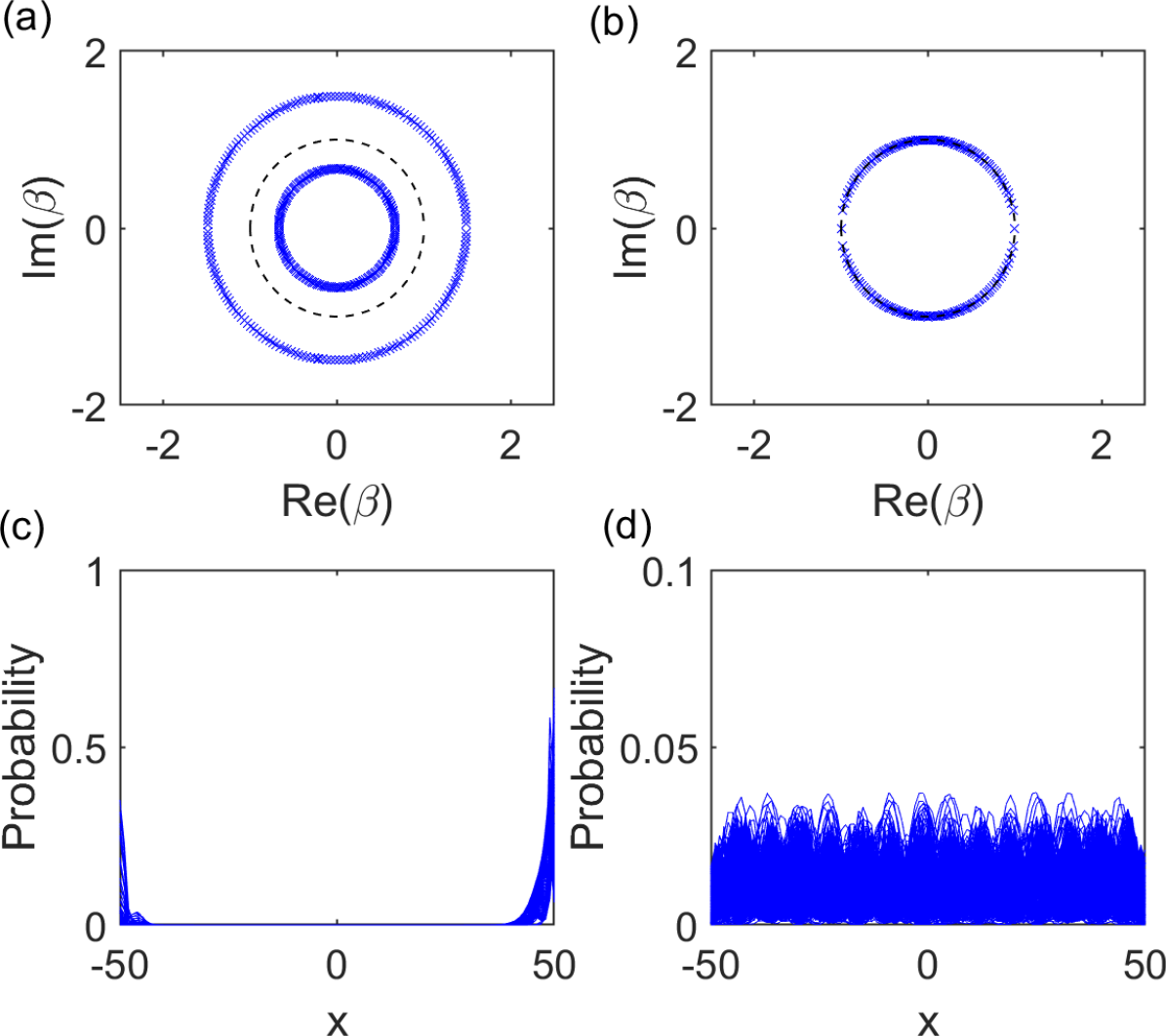}
  \caption{GBZ and BZ of the time evolution operator $U$ under RBC in the case of $\gamma=0.5$, $\theta_1=3\pi/10$ and $\theta_3=\pi/4$ for (a) $\theta_2=\pi/5$, (b) $\theta_2=\pi/6$. GBZ is depicted by blue crosses while BZ forming a unit circle on the complex plane is represented by dashed line. (c,d) The probability distributions of the corresponding eigenvectors of $U$ for (c) $\theta_2=\pi/2$ and (b) $\theta_2=\pi/6$.}
  \label{fig:GBZ}
\end{figure}

Here, we check the bulk-edge correspondence for $\mathbb{Z}_2$ point-gap topology of our quantum walk.
As explained, in case of Fig.\ \ref{fig:spectra} (a), the point gap opens in which $\mathbb{Z}_2$ topological invariant is nontrivial, while there are no point gap in Fig.\ \ref{fig:spectra} (b).
To calculate the spectra and eigenstates of the operator $U$ with RBC in the real space representation, we apply for the the generalized Brillouin zone (GBZ) \cite{PhysRevLett.121.086803,PhysRevLett.123.066404,PhysRevLett.125.226402,PhysRevB.101.195147} as shown in Fig.\ \ref{fig:GBZ} (a) and (b).  Figure \ref{fig:spectra} (a) shows that the spectrum for the system with RBC (RBC spectrum, in short) depicted by blue dots appears as arcs in the point gap. Furthermore, the eigenstates are localized at right and left boundaries, confirming $\mathbb{Z}_2$ NHSE. More specifically, as shown in Fig.\ \ref{fig:GBZ} (a) and (c), GBZ inside of Brillouin zone (BZ) correspond to eigenmodes localized at the right boundary, while those outside of Brillouin zone (BZ) represent left-boundary-localized eigenmodes \cite{PhysRevLett.125.126402}.
On the other hand, in case of Fig.\ \ref{fig:spectra}(b) with trivial topology, the RBC spectrum almost overlaps with the PBC spectrum. Moreover, the GBZ coincides with the BZ as shown in Fig.\ \ref{fig:GBZ}(b), resulting in the disappearance of NHSE, and the eigenmodes are not localized near the boundaries as shown in Fig. \ref{fig:GBZ} (d).
These numerical results verify that the quantum walk in Eq.\ (\ref{eq:time evolution operator}) exhibits $\mathbb{Z}_2$ point-gap topology in $H$ type and satisfies the bulk-edge correspondence for $\mathbb{Z}_2$ point-gap topology.

\begin{figure}[tb]
  \centering
  \includegraphics[width=\columnwidth]{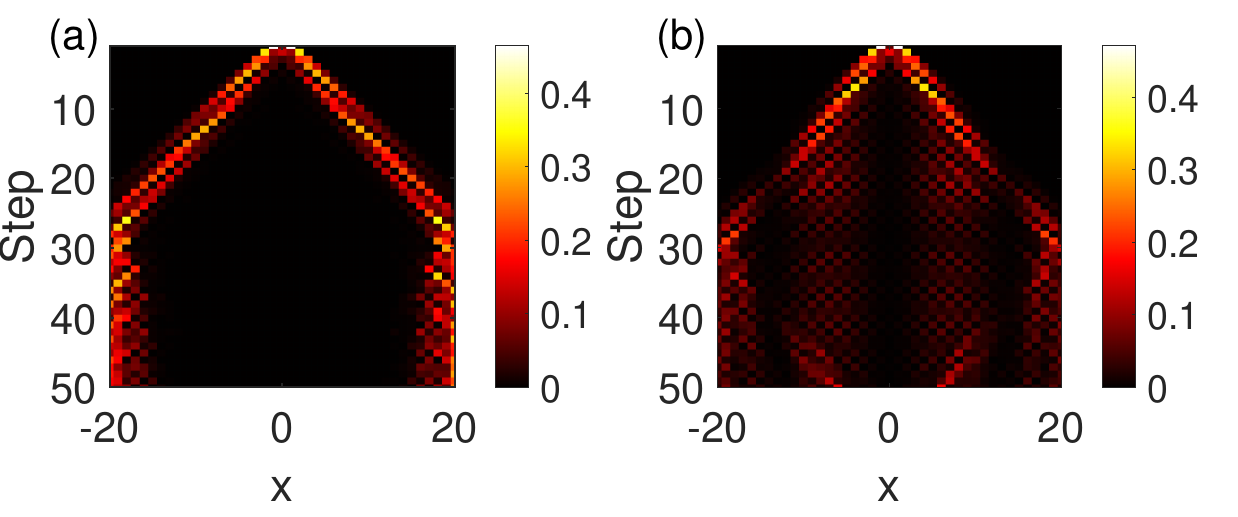}
  \caption{Time evolution of the probablility distribution of the quantum walk in the case of $\gamma=0.5$, $\theta_1=3\pi/10$ and $\theta_3=\pi/4$. (a) $\theta_2=\pi/5$, (b) $\theta_2=\pi/6$. The initial state of the walker is given by $|\psi_0\rangle=|0\rangle \otimes \left[|a\rangle \otimes (|L\rangle+i|R\rangle)+|b\rangle \otimes (|L\rangle-i|R\rangle)\right]/2$.}
  \label{fig:evolution_qw}
\end{figure}

Finally, we consider dynamics of this quantum walk.
We consider a walker initially positioned at the origin and investigate its descrete evolution under the presence and absence of the skin effect, respectively. In Fig.\ \ref{fig:evolution_qw}, we present the probability distribution of the walker in real space after each step. It can be observed that in systems where the time evolution operator exhibits a point gap, the walker becomes localized near the left and right boundaries after a certain number of steps. In contrast, in systems where the time evolution operator is topologically trivial, the walker is not confined to the boundaries, indicating the absence of the skin effect.

\subsection{$\mathbb{Z}_2$ point-gap topology in junction systems}
\label{subsec:junction system}

\begin{figure}[tb]
  \centering
  \includegraphics[width=\columnwidth]{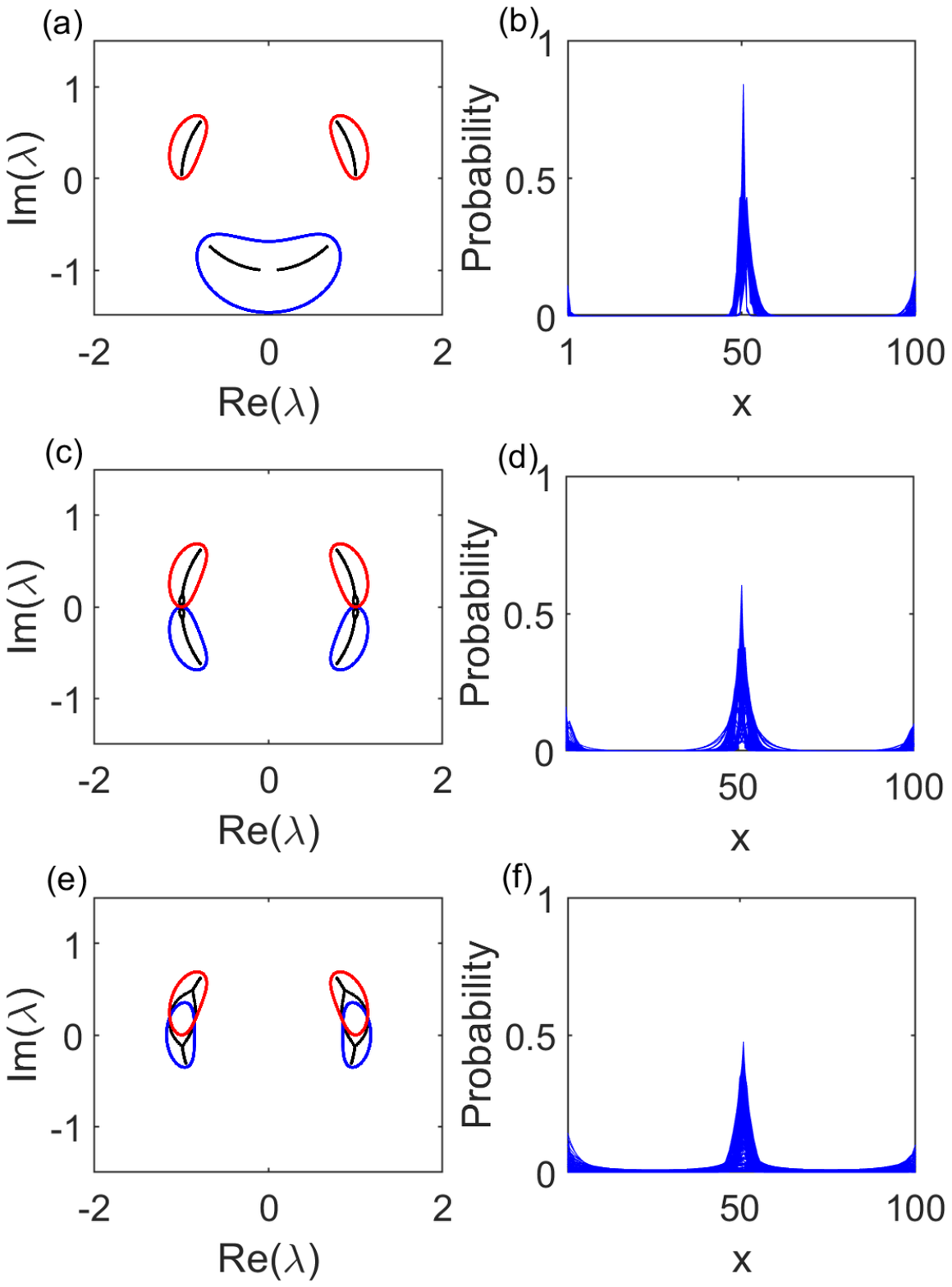}
  \caption{(a,c,e) Eigenvalue spectra and (b,d,f) probability distributions of all eigenvectors of the time-evolution operator in the junction system composed of two non-trivial subsystems $A$ and $B$ with the parameters $l=50$, $\gamma_A=\gamma_B=0.5$, $\theta_{1,B}=3\pi/10$, $\theta_{2A}=\theta_{2B}=\pi/4$ and $\theta_{3A}=\theta_{3B}=2\pi/5$, (a,b) $\theta_{1A}=\pi/10$, (c,d) $\theta_{1A}=\pi/5$, (e,f) $\theta_{1A}=\pi/4$. The left figures (a,c,e) show the eigenvalue spectrum of subsystem A (blue curve), subsystem B (red curve) and junction system with PBC (black curve), while the right figures (b,d,f) show the corresponding distribution of all eigenstates.}
  \label{fig:junction system}
\end{figure}

\begin{figure}[tb]
  \centering
  \includegraphics[width=\columnwidth]{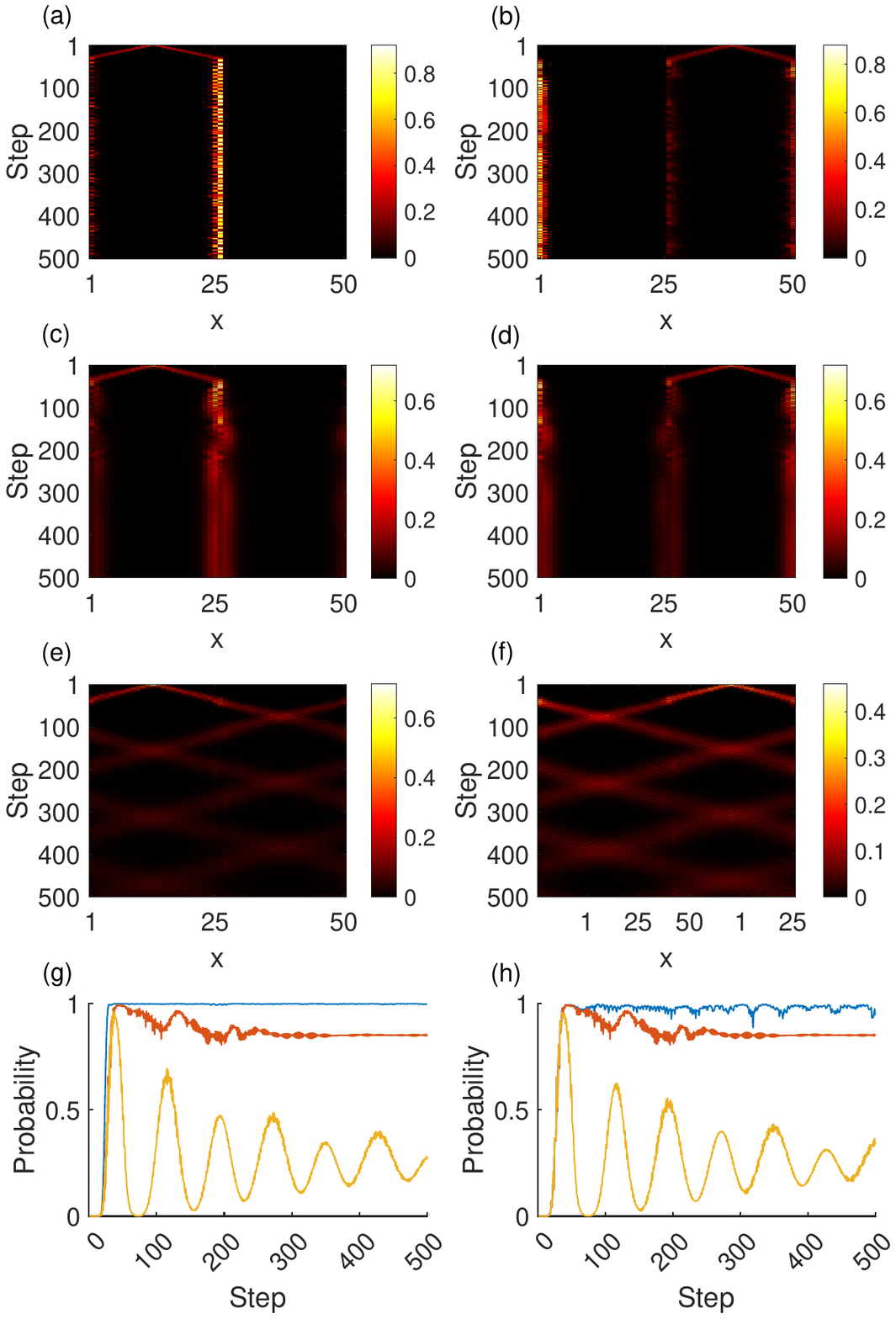}
  \caption{(a-f) Time evolution of the probability distribution in a junction system with PBC with the parameters $l=25$, $\gamma_A=\gamma_B=0.5$, $\theta_{1,B}=3\pi/10$, $\theta_{2A}=\theta_{2B}=\pi/4$ and $\theta_{3A}=\theta_{3B}=2\pi
/5$, (a,b) $\theta_{1A}=\pi/10$, (c,d) $\theta_{1A}=\pi/5$, (e,f) $\theta_{1A}=\pi/4$. (g,h) The time dependence of the probability of the walker near the boundaries of two subsystems. 
The initinal state of the walker is given by (a,c,e,g) $|\psi_0\rangle=|x_{0,A}\rangle \otimes\left[|a\rangle \otimes (|L\rangle+i|R\rangle)+|b\rangle \otimes (|L\rangle-i|R\rangle)\right]/2$ and (b,d,f,h) $|\psi_0\rangle=|x_{0,B}\rangle \otimes\left[|a\rangle \otimes (|L\rangle+i|R\rangle)+|b\rangle \otimes(|L\rangle-i|R\rangle)\right]/2$, with $x_{0,A}=(l+1)/2$ and $x_{0,B}=l+(l+1)/2$.}
  \label{fig:evo_junc}
\end{figure}
 
Recently, the bulk-edge correspondence for $\mathbb{Z}$ point-gap topology has been generalized to junction systems where the topological number varies at interfaces\cite{PhysRevB.108.L121302}. It states that the spectrum for the junction systems in a ring geometry appears in a region where the difference in the point-gap topological numbers for two subsystems is nonzero and the corresponding eigenstates are localized at the interfaces. 

In this subsection, we study the bulk-edge correspondence for $\mathbb{Z}_2$  point gap topology in junction systems by using the non-unitary quantum walk with $\mathbb{Z}_2$ point-gap topology. We explore a system with a ring geometry comprising two subsystems connected; subsystem $A$ in the region $1\leq x\leq l$ and subsystem $B$ in the region $l+1\leq x\leq 2l$, where both subsystems possess distinct parameters of the non-unitary quantum walk with $\mathbb{Z}_2$ point-gap topology. Therefore, there are two interfaces near $x=1$ and $l$. 

Figure. \ref{fig:junction system} (a) shows PBC spectra for the individual subsystems, $\sigma^A_\text{PBC}$ and $\sigma^B_\text{PBC}$, under different parameter pairs ($\theta_1^A$, $\theta_1^B$), as well as the eigenvalue spectrum of the junction system, $\sigma^{\text{junc}}_\text{PBC}$. Note that the $\mathbb{Z}_2$ topological number for each subsystem $\nu$ is one in the closed curves of $\sigma^A_\text{PBC}$ or $\sigma^B_\text{PBC}$. Thereby, the difference in $\mathbb{Z}_2$ topological topological number is zero in a region with no point gap and a region where $\sigma^A_\text{PBC}$ and $\sigma^B_\text{PBC}$ are overlapped.
We observed that $\sigma^{\text{junc}}_\text{PBC}$ appears as arcs in the region where the difference in $\mathbb{Z}_2$ topological topological number is nonzero as shown by black curves in Fig.\ \ref{fig:junction system} (a). 
Furthermore, the eigenstates localized near the interfaces of the two subsystems, called the non-Hermitian proximity effects\cite{PhysRevB.108.L121302}, as shown in Fig. \ref{fig:junction system}(b). 
We note that although there is always a tendency for the eigenstates of the time evolution operator to be localized near the boundaries of the two subsystems, this localization strength gradually weakens as $\sigma^A_\text{PBC}$ and $\sigma^B_\text{PBC}$ get closer. When $\sigma^A_\text{PBC}$ and $\sigma^B_\text{PBC}$ are sufficiently far apart, the eigenstates are tightly confined near the interface [see Fig.\ref{fig:junction system}(a) and (b)]. However, as the distance between the two spectra decreases and even when they intersect, the eigenstates may have a non-negligible probability of appearing even at the center of the subsystems, although they still exhibit a tendency to move towards the interface [see Fig.\ref{fig:junction system}(e) and (f)].
These observations establishes the bulk-edge correspondence for $\mathbb{Z}_2$ point-gap topology in junction systems in the same fashion with $\mathbb{Z}$ point-gap topology as clarified in Ref. \cite{PhysRevB.108.L121302}.

Figure \ref{fig:evo_junc} illustrates the time evolution of the walker in the junction system. In Figs. \ref{fig:evo_junc}(a), (c),(e), and (g), the walker is initially located on subsystem $A$, while in Figs. \ref{fig:evo_junc}(b), (d),(f), and (h), the walker's initial state is on subsystem $B$. We observe that when $\sigma^A_\text{PBC}$ and $\sigma^B_\text{PBC}$ are sufficiently far apart, the walker is strongly stucked near the interface. 
However, in cases where $\sigma^A_\text{PBC}$ and $\sigma^B_\text{PBC}$ overlap each other, the confinement near the interface of the subsystems is not as strong as in the previous cases, as indicated by Figs. \ref{fig:evo_junc}(g,h).

\section{Summary}
\label{sec:summary}
In summary, in the present work, first we have proposed the dual symmetry classification based on the fact that there is no strict difference in mathematical definition between non-Hermitian Hamiltonians and non-unitary time-evolution operators. By using this dual symmetry classification and regarding the time-evolution operator as the non-Hermitian Hamiltonian, we have successfully constructed a non-unitary quantum walk with spinful $\text{TRS}^{\dagger}$. Furthermore, we have shown that this non-unitary quantum walk can obtain non-trivial $\mathbb{Z}_2$ point-gap topology again regarding the time-evolution operator as the non-Hermitian Hamiltonian. The eigenvalue spectrum of the system strongly depends on the system's boundary conditions when the time evolution operator is $\mathbb{Z}_2$ topologically non-trivial. Under RBC, the eigenstates of the time evolution operator exhibit localization near the boundaries, indicating the presence of the skin effect. In contrast to the traditional Hatano-Nelson model, in such a quantum walk, we can control the opening and closing of the point gap by adjusting the parameters of the coin operators or the gain-loss operator. 

Moreover, we studied quantum walks in junction system composed of two $\mathbb{Z}_2$ non-trivial quantum walks. When the distance between the PBC spectra of the two topologically non-trivial subsystems in the junction system is sufficiently large, the PBC spectrum of the junction system can be effectively treated as an RBC spectrum, which is also a novel approach to achieve RBC in arbitrary quantum walks. In this scenario, all eigenstates become localized near the boundaries of the two subsystems, resulting in the non-Hermitian proximity effects. However, as the distance between the two PBC spectra decreases, the confinement ability of the boundaries on the eigenstates weakens. The same phenomenon also occurs with the confinement ability on the walker.

In this work, we have defined the skin effect in non-unitary quantum walk systems as the localization of eigenstates of the time evolution operator near the boundaries of the system. We have noticed that in quantum walk systems with skin effect, the walker becomes localized near the boundaries of the system after a sufficiently long time. However, especially for the junction system, the state of the walker's evolution over time is not well-defined. In particular, when the eigenvalue spectrum of the PBC junction system has a non-zero area, the choice of initial state can affect the boundaries to which the walker becomes confined. Studying the time evolution of non-unitary quantum walks is an interesting work and holds significant importance for investigating the time-dependent topological properties of non-Hermitian systems.

\added{Finally, we discuss a possible application of the dual symmetry classification proposed in the current work to unitary time-evolution operators. In principle, the dual symmetry classification can also be applied to unitary time-evolution systems. Applying the $H$-type classification, we can identify line and point gaps for various unitary time-evolution operators, where non-trivial topological phases may arise. However, these unitary time-evolution operators do not show phenomena peculiar to non-Hermitian physics, such as point-gap topology and skin effects, as long as they keep unitarity, for the following reasons. A unitary time-evolution operator $U$, related to the effective Hermitian $H_{\text{eff}}$ by Eq.\ (\ref{eq:H_eff}), can commute with $H_{\text{eff}}$, leading to $U$ and $H_{\text{eff}}$ having simultaneous eigenstates. Therefore, since all the eigenstates are those of the Hermitian Hamiltonian, phenomena unique to non-Hermitian systems are not anticipated.
However, applying the dual symmetry classification for unitary time-evolution operators may give an opportunity to better understand of the topological properties in Hermitian systems. We leave it as an open problem.}

\section*{Acknowledgements}

We thank Yasuhiro\ Asano, Geonhwi Hwang, \added{Franco Nori,} Masatoshi\ Sato, and Kousuke\ Yakubo for helpful discussions. Zhiyu Jiang was supported by JST SPRING (Grant No. JPMJSP2119). This work was supported by JSPS KAKENHI (Grants No. JP20H01828, No. JP22K03463, No. JP22H01140, and JP24K00545).


\bibliographystyle{apsrev4-2}
\bibliography{Z2point-gap_QW_20240513}

\begin{thebibliography}{67}%
\makeatletter
\providecommand \@ifxundefined [1]{%
 \@ifx{#1\undefined}
}%
\providecommand \@ifnum [1]{%
 \ifnum #1\expandafter \@firstoftwo
 \else \expandafter \@secondoftwo
 \fi
}%
\providecommand \@ifx [1]{%
 \ifx #1\expandafter \@firstoftwo
 \else \expandafter \@secondoftwo
 \fi
}%
\providecommand \natexlab [1]{#1}%
\providecommand \enquote  [1]{``#1''}%
\providecommand \bibnamefont  [1]{#1}%
\providecommand \bibfnamefont [1]{#1}%
\providecommand \citenamefont [1]{#1}%
\providecommand \href@noop [0]{\@secondoftwo}%
\providecommand \href [0]{\begingroup \@sanitize@url \@href}%
\providecommand \@href[1]{\@@startlink{#1}\@@href}%
\providecommand \@@href[1]{\endgroup#1\@@endlink}%
\providecommand \@sanitize@url [0]{\catcode `\\12\catcode `\$12\catcode
  `\&12\catcode `\#12\catcode `\^12\catcode `\_12\catcode `\%12\relax}%
\providecommand \@@startlink[1]{}%
\providecommand \@@endlink[0]{}%
\providecommand \url  [0]{\begingroup\@sanitize@url \@url }%
\providecommand \@url [1]{\endgroup\@href {#1}{\urlprefix }}%
\providecommand \urlprefix  [0]{URL }%
\providecommand \Eprint [0]{\href }%
\providecommand \doibase [0]{https://doi.org/}%
\providecommand \selectlanguage [0]{\@gobble}%
\providecommand \bibinfo  [0]{\@secondoftwo}%
\providecommand \bibfield  [0]{\@secondoftwo}%
\providecommand \translation [1]{[#1]}%
\providecommand \BibitemOpen [0]{}%
\providecommand \bibitemStop [0]{}%
\providecommand \bibitemNoStop [0]{.\EOS\space}%
\providecommand \EOS [0]{\spacefactor3000\relax}%
\providecommand \BibitemShut  [1]{\csname bibitem#1\endcsname}%
\let\auto@bib@innerbib\@empty
\bibitem [{\citenamefont {Yuto~Ashida}\ and\ \citenamefont
  {Ueda}(2020)}]{Ashida-Review}%
  \BibitemOpen
  \bibfield  {author} {\bibinfo {author} {\bibfnamefont {Z.~G.}\ \bibnamefont
  {Yuto~Ashida}}\ and\ \bibinfo {author} {\bibfnamefont {M.}~\bibnamefont
  {Ueda}},\ }\href {https://doi.org/10.1080/00018732.2021.1876991} {\bibfield
  {journal} {\bibinfo  {journal} {Advances in Physics}\ }\textbf {\bibinfo
  {volume} {69}},\ \bibinfo {pages} {249} (\bibinfo {year} {2020})}\BibitemShut
  {NoStop}%
\bibitem [{\citenamefont {Okuma}\ and\ \citenamefont
  {Sato}(2023)}]{Okuma-Review}%
  \BibitemOpen
  \bibfield  {author} {\bibinfo {author} {\bibfnamefont {N.}~\bibnamefont
  {Okuma}}\ and\ \bibinfo {author} {\bibfnamefont {M.}~\bibnamefont {Sato}},\
  }\href
  {https://doi.org/https://doi.org/10.1146/annurev-conmatphys-040521-033133}
  {\bibfield  {journal} {\bibinfo  {journal} {Annual Review of Condensed Matter
  Physics}\ }\textbf {\bibinfo {volume} {14}},\ \bibinfo {pages} {83} (\bibinfo
  {year} {2023})}\BibitemShut {NoStop}%
\bibitem [{\citenamefont {Lin}\ \emph {et~al.}(2011)\citenamefont {Lin},
  \citenamefont {Ramezani}, \citenamefont {Eichelkraut}, \citenamefont
  {Kottos}, \citenamefont {Cao},\ and\ \citenamefont
  {Christodoulides}}]{PhysRevLett.106.213901}%
  \BibitemOpen
  \bibfield  {author} {\bibinfo {author} {\bibfnamefont {Z.}~\bibnamefont
  {Lin}}, \bibinfo {author} {\bibfnamefont {H.}~\bibnamefont {Ramezani}},
  \bibinfo {author} {\bibfnamefont {T.}~\bibnamefont {Eichelkraut}}, \bibinfo
  {author} {\bibfnamefont {T.}~\bibnamefont {Kottos}}, \bibinfo {author}
  {\bibfnamefont {H.}~\bibnamefont {Cao}},\ and\ \bibinfo {author}
  {\bibfnamefont {D.~N.}\ \bibnamefont {Christodoulides}},\ }\href
  {https://doi.org/10.1103/PhysRevLett.106.213901} {\bibfield  {journal}
  {\bibinfo  {journal} {Phys. Rev. Lett.}\ }\textbf {\bibinfo {volume} {106}},\
  \bibinfo {pages} {213901} (\bibinfo {year} {2011})}\BibitemShut {NoStop}%
\bibitem [{\citenamefont {Regensburger}\ \emph {et~al.}(2012)\citenamefont
  {Regensburger}, \citenamefont {Bersch}, \citenamefont {Miri}, \citenamefont
  {Onishchukov}, \citenamefont {Christodoulides},\ and\ \citenamefont
  {Peschel}}]{regensburger2012parity}%
  \BibitemOpen
  \bibfield  {author} {\bibinfo {author} {\bibfnamefont {A.}~\bibnamefont
  {Regensburger}}, \bibinfo {author} {\bibfnamefont {C.}~\bibnamefont
  {Bersch}}, \bibinfo {author} {\bibfnamefont {M.-A.}\ \bibnamefont {Miri}},
  \bibinfo {author} {\bibfnamefont {G.}~\bibnamefont {Onishchukov}}, \bibinfo
  {author} {\bibfnamefont {D.~N.}\ \bibnamefont {Christodoulides}},\ and\
  \bibinfo {author} {\bibfnamefont {U.}~\bibnamefont {Peschel}},\ }\href
  {https://www.nature.com/articles/nature11298} {\bibfield  {journal} {\bibinfo
   {journal} {Nature}\ }\textbf {\bibinfo {volume} {488}},\ \bibinfo {pages}
  {167} (\bibinfo {year} {2012})}\BibitemShut {NoStop}%
\bibitem [{\citenamefont {Feng}\ \emph {et~al.}(2013)\citenamefont {Feng},
  \citenamefont {Xu}, \citenamefont {Fegadolli}, \citenamefont {Lu},
  \citenamefont {Oliveira}, \citenamefont {Almeida}, \citenamefont {Chen},\
  and\ \citenamefont {Scherer}}]{feng2013experimental}%
  \BibitemOpen
  \bibfield  {author} {\bibinfo {author} {\bibfnamefont {L.}~\bibnamefont
  {Feng}}, \bibinfo {author} {\bibfnamefont {Y.-L.}\ \bibnamefont {Xu}},
  \bibinfo {author} {\bibfnamefont {W.~S.}\ \bibnamefont {Fegadolli}}, \bibinfo
  {author} {\bibfnamefont {M.-H.}\ \bibnamefont {Lu}}, \bibinfo {author}
  {\bibfnamefont {J.~E.}\ \bibnamefont {Oliveira}}, \bibinfo {author}
  {\bibfnamefont {V.~R.}\ \bibnamefont {Almeida}}, \bibinfo {author}
  {\bibfnamefont {Y.-F.}\ \bibnamefont {Chen}},\ and\ \bibinfo {author}
  {\bibfnamefont {A.}~\bibnamefont {Scherer}},\ }\href
  {https://www.nature.com/articles/nmat3495} {\bibfield  {journal} {\bibinfo
  {journal} {Nature materials}\ }\textbf {\bibinfo {volume} {12}},\ \bibinfo
  {pages} {108} (\bibinfo {year} {2013})}\BibitemShut {NoStop}%
\bibitem [{\citenamefont {Weimann}\ \emph {et~al.}(2017)\citenamefont
  {Weimann}, \citenamefont {Kremer}, \citenamefont {Plotnik}, \citenamefont
  {Lumer}, \citenamefont {Nolte}, \citenamefont {Makris}, \citenamefont
  {Segev}, \citenamefont {Rechtsman},\ and\ \citenamefont
  {Szameit}}]{weimann2017topologically}%
  \BibitemOpen
  \bibfield  {author} {\bibinfo {author} {\bibfnamefont {S.}~\bibnamefont
  {Weimann}}, \bibinfo {author} {\bibfnamefont {M.}~\bibnamefont {Kremer}},
  \bibinfo {author} {\bibfnamefont {Y.}~\bibnamefont {Plotnik}}, \bibinfo
  {author} {\bibfnamefont {Y.}~\bibnamefont {Lumer}}, \bibinfo {author}
  {\bibfnamefont {S.}~\bibnamefont {Nolte}}, \bibinfo {author} {\bibfnamefont
  {K.~G.}\ \bibnamefont {Makris}}, \bibinfo {author} {\bibfnamefont
  {M.}~\bibnamefont {Segev}}, \bibinfo {author} {\bibfnamefont {M.~C.}\
  \bibnamefont {Rechtsman}},\ and\ \bibinfo {author} {\bibfnamefont
  {A.}~\bibnamefont {Szameit}},\ }\href
  {https://www.nature.com/articles/nmat4811} {\bibfield  {journal} {\bibinfo
  {journal} {Nature materials}\ }\textbf {\bibinfo {volume} {16}},\ \bibinfo
  {pages} {433} (\bibinfo {year} {2017})}\BibitemShut {NoStop}%
\bibitem [{\citenamefont {Zhou}\ \emph {et~al.}(2018)\citenamefont {Zhou},
  \citenamefont {Peng}, \citenamefont {Yoon}, \citenamefont {Hsu},
  \citenamefont {Nelson}, \citenamefont {Fu}, \citenamefont {Joannopoulos},
  \citenamefont {Solja{\v{c}}i{\'c}},\ and\ \citenamefont
  {Zhen}}]{zhou2018observation}%
  \BibitemOpen
  \bibfield  {author} {\bibinfo {author} {\bibfnamefont {H.}~\bibnamefont
  {Zhou}}, \bibinfo {author} {\bibfnamefont {C.}~\bibnamefont {Peng}}, \bibinfo
  {author} {\bibfnamefont {Y.}~\bibnamefont {Yoon}}, \bibinfo {author}
  {\bibfnamefont {C.~W.}\ \bibnamefont {Hsu}}, \bibinfo {author} {\bibfnamefont
  {K.~A.}\ \bibnamefont {Nelson}}, \bibinfo {author} {\bibfnamefont
  {L.}~\bibnamefont {Fu}}, \bibinfo {author} {\bibfnamefont {J.~D.}\
  \bibnamefont {Joannopoulos}}, \bibinfo {author} {\bibfnamefont
  {M.}~\bibnamefont {Solja{\v{c}}i{\'c}}},\ and\ \bibinfo {author}
  {\bibfnamefont {B.}~\bibnamefont {Zhen}},\ }\href
  {https://www.science.org/doi/full/10.1126/science.aap9859} {\bibfield
  {journal} {\bibinfo  {journal} {Science}\ }\textbf {\bibinfo {volume}
  {359}},\ \bibinfo {pages} {1009} (\bibinfo {year} {2018})}\BibitemShut
  {NoStop}%
\bibitem [{\citenamefont {Zhao}\ \emph {et~al.}(2019)\citenamefont {Zhao},
  \citenamefont {Qiao}, \citenamefont {Wu}, \citenamefont {Midya},
  \citenamefont {Longhi},\ and\ \citenamefont
  {Feng}}]{doi:10.1126/science.aay1064}%
  \BibitemOpen
  \bibfield  {author} {\bibinfo {author} {\bibfnamefont {H.}~\bibnamefont
  {Zhao}}, \bibinfo {author} {\bibfnamefont {X.}~\bibnamefont {Qiao}}, \bibinfo
  {author} {\bibfnamefont {T.}~\bibnamefont {Wu}}, \bibinfo {author}
  {\bibfnamefont {B.}~\bibnamefont {Midya}}, \bibinfo {author} {\bibfnamefont
  {S.}~\bibnamefont {Longhi}},\ and\ \bibinfo {author} {\bibfnamefont
  {L.}~\bibnamefont {Feng}},\ }\href {https://doi.org/10.1126/science.aay1064}
  {\bibfield  {journal} {\bibinfo  {journal} {Science}\ }\textbf {\bibinfo
  {volume} {365}},\ \bibinfo {pages} {1163} (\bibinfo {year}
  {2019})}\BibitemShut {NoStop}%
\bibitem [{\citenamefont {Ningyuan}\ \emph {et~al.}(2015)\citenamefont
  {Ningyuan}, \citenamefont {Owens}, \citenamefont {Sommer}, \citenamefont
  {Schuster},\ and\ \citenamefont {Simon}}]{PhysRevX.5.021031}%
  \BibitemOpen
  \bibfield  {author} {\bibinfo {author} {\bibfnamefont {J.}~\bibnamefont
  {Ningyuan}}, \bibinfo {author} {\bibfnamefont {C.}~\bibnamefont {Owens}},
  \bibinfo {author} {\bibfnamefont {A.}~\bibnamefont {Sommer}}, \bibinfo
  {author} {\bibfnamefont {D.}~\bibnamefont {Schuster}},\ and\ \bibinfo
  {author} {\bibfnamefont {J.}~\bibnamefont {Simon}},\ }\href
  {https://doi.org/10.1103/PhysRevX.5.021031} {\bibfield  {journal} {\bibinfo
  {journal} {Phys. Rev. X}\ }\textbf {\bibinfo {volume} {5}},\ \bibinfo {pages}
  {021031} (\bibinfo {year} {2015})}\BibitemShut {NoStop}%
\bibitem [{\citenamefont {Ezawa}(2019{\natexlab{a}})}]{PhysRevB.99.201411}%
  \BibitemOpen
  \bibfield  {author} {\bibinfo {author} {\bibfnamefont {M.}~\bibnamefont
  {Ezawa}},\ }\href {https://doi.org/10.1103/PhysRevB.99.201411(R)} {\bibfield
  {journal} {\bibinfo  {journal} {Phys. Rev. B}\ }\textbf {\bibinfo {volume}
  {99}},\ \bibinfo {pages} {201411} (\bibinfo {year}
  {2019}{\natexlab{a}})}\BibitemShut {NoStop}%
\bibitem [{\citenamefont {Ezawa}(2019{\natexlab{b}})}]{PhysRevB.100.045407}%
  \BibitemOpen
  \bibfield  {author} {\bibinfo {author} {\bibfnamefont {M.}~\bibnamefont
  {Ezawa}},\ }\href {https://doi.org/10.1103/PhysRevB.100.045407} {\bibfield
  {journal} {\bibinfo  {journal} {Phys. Rev. B}\ }\textbf {\bibinfo {volume}
  {100}},\ \bibinfo {pages} {045407} (\bibinfo {year}
  {2019}{\natexlab{b}})}\BibitemShut {NoStop}%
\bibitem [{\citenamefont {Helbig}\ \emph {et~al.}(2020)\citenamefont {Helbig},
  \citenamefont {Hofmann}, \citenamefont {Imhof}, \citenamefont {Abdelghany},
  \citenamefont {Kiessling}, \citenamefont {Molenkamp}, \citenamefont {Lee},
  \citenamefont {Szameit}, \citenamefont {Greiter},\ and\ \citenamefont
  {Thomale}}]{helbig2020generalized}%
  \BibitemOpen
  \bibfield  {author} {\bibinfo {author} {\bibfnamefont {T.}~\bibnamefont
  {Helbig}}, \bibinfo {author} {\bibfnamefont {T.}~\bibnamefont {Hofmann}},
  \bibinfo {author} {\bibfnamefont {S.}~\bibnamefont {Imhof}}, \bibinfo
  {author} {\bibfnamefont {M.}~\bibnamefont {Abdelghany}}, \bibinfo {author}
  {\bibfnamefont {T.}~\bibnamefont {Kiessling}}, \bibinfo {author}
  {\bibfnamefont {L.}~\bibnamefont {Molenkamp}}, \bibinfo {author}
  {\bibfnamefont {C.}~\bibnamefont {Lee}}, \bibinfo {author} {\bibfnamefont
  {A.}~\bibnamefont {Szameit}}, \bibinfo {author} {\bibfnamefont
  {M.}~\bibnamefont {Greiter}},\ and\ \bibinfo {author} {\bibfnamefont
  {R.}~\bibnamefont {Thomale}},\ }\href
  {https://www.nature.com/articles/s41567-020-0922-9} {\bibfield  {journal}
  {\bibinfo  {journal} {Nature Physics}\ }\textbf {\bibinfo {volume} {16}},\
  \bibinfo {pages} {747} (\bibinfo {year} {2020})}\BibitemShut {NoStop}%
\bibitem [{\citenamefont {Yoshida}\ \emph {et~al.}(2020)\citenamefont
  {Yoshida}, \citenamefont {Mizoguchi},\ and\ \citenamefont
  {Hatsugai}}]{PhysRevResearch.2.022062}%
  \BibitemOpen
  \bibfield  {author} {\bibinfo {author} {\bibfnamefont {T.}~\bibnamefont
  {Yoshida}}, \bibinfo {author} {\bibfnamefont {T.}~\bibnamefont {Mizoguchi}},\
  and\ \bibinfo {author} {\bibfnamefont {Y.}~\bibnamefont {Hatsugai}},\ }\href
  {https://doi.org/10.1103/PhysRevResearch.2.022062} {\bibfield  {journal}
  {\bibinfo  {journal} {Phys. Rev. Res.}\ }\textbf {\bibinfo {volume} {2}},\
  \bibinfo {pages} {022062(R)} (\bibinfo {year} {2020})}\BibitemShut {NoStop}%
\bibitem [{\citenamefont {Hofmann}\ \emph {et~al.}(2020)\citenamefont
  {Hofmann}, \citenamefont {Helbig}, \citenamefont {Schindler}, \citenamefont
  {Salgo}, \citenamefont {Brzezi\ifmmode~\acute{n}\else \'{n}\fi{}ska},
  \citenamefont {Greiter}, \citenamefont {Kiessling}, \citenamefont {Wolf},
  \citenamefont {Vollhardt}, \citenamefont {Kaba\ifmmode~\check{s}\else
  \v{s}\fi{}i}, \citenamefont {Lee}, \citenamefont {Bilu\ifmmode \check{s}\else
  \v{s}\fi{}i\ifmmode~\acute{c}\else \'{c}\fi{}}, \citenamefont {Thomale},\
  and\ \citenamefont {Neupert}}]{PhysRevResearch.2.023265}%
  \BibitemOpen
  \bibfield  {author} {\bibinfo {author} {\bibfnamefont {T.}~\bibnamefont
  {Hofmann}}, \bibinfo {author} {\bibfnamefont {T.}~\bibnamefont {Helbig}},
  \bibinfo {author} {\bibfnamefont {F.}~\bibnamefont {Schindler}}, \bibinfo
  {author} {\bibfnamefont {N.}~\bibnamefont {Salgo}}, \bibinfo {author}
  {\bibfnamefont {M.}~\bibnamefont {Brzezi\ifmmode~\acute{n}\else
  \'{n}\fi{}ska}}, \bibinfo {author} {\bibfnamefont {M.}~\bibnamefont
  {Greiter}}, \bibinfo {author} {\bibfnamefont {T.}~\bibnamefont {Kiessling}},
  \bibinfo {author} {\bibfnamefont {D.}~\bibnamefont {Wolf}}, \bibinfo {author}
  {\bibfnamefont {A.}~\bibnamefont {Vollhardt}}, \bibinfo {author}
  {\bibfnamefont {A.}~\bibnamefont {Kaba\ifmmode~\check{s}\else \v{s}\fi{}i}},
  \bibinfo {author} {\bibfnamefont {C.~H.}\ \bibnamefont {Lee}}, \bibinfo
  {author} {\bibfnamefont {A.}~\bibnamefont {Bilu\ifmmode \check{s}\else
  \v{s}\fi{}i\ifmmode~\acute{c}\else \'{c}\fi{}}}, \bibinfo {author}
  {\bibfnamefont {R.}~\bibnamefont {Thomale}},\ and\ \bibinfo {author}
  {\bibfnamefont {T.}~\bibnamefont {Neupert}},\ }\href
  {https://doi.org/10.1103/PhysRevResearch.2.023265} {\bibfield  {journal}
  {\bibinfo  {journal} {Phys. Rev. Res.}\ }\textbf {\bibinfo {volume} {2}},\
  \bibinfo {pages} {023265} (\bibinfo {year} {2020})}\BibitemShut {NoStop}%
\bibitem [{\citenamefont {Liu}\ \emph {et~al.}(2021)\citenamefont {Liu},
  \citenamefont {Shao}, \citenamefont {Ma}, \citenamefont {Zhang},
  \citenamefont {You}, \citenamefont {Wu}, \citenamefont {Xiang}, \citenamefont
  {Cui},\ and\ \citenamefont {Zhang}}]{liu2021non}%
  \BibitemOpen
  \bibfield  {author} {\bibinfo {author} {\bibfnamefont {S.}~\bibnamefont
  {Liu}}, \bibinfo {author} {\bibfnamefont {R.}~\bibnamefont {Shao}}, \bibinfo
  {author} {\bibfnamefont {S.}~\bibnamefont {Ma}}, \bibinfo {author}
  {\bibfnamefont {L.}~\bibnamefont {Zhang}}, \bibinfo {author} {\bibfnamefont
  {O.}~\bibnamefont {You}}, \bibinfo {author} {\bibfnamefont {H.}~\bibnamefont
  {Wu}}, \bibinfo {author} {\bibfnamefont {Y.~J.}\ \bibnamefont {Xiang}},
  \bibinfo {author} {\bibfnamefont {T.~J.}\ \bibnamefont {Cui}},\ and\ \bibinfo
  {author} {\bibfnamefont {S.}~\bibnamefont {Zhang}},\ }\href
  {https://spj.science.org/doi/full/10.34133/2021/5608038} {\bibfield
  {journal} {\bibinfo  {journal} {Research}\ } (\bibinfo {year}
  {2021})}\BibitemShut {NoStop}%
\bibitem [{\citenamefont {Diehl}\ \emph {et~al.}(2011)\citenamefont {Diehl},
  \citenamefont {Rico}, \citenamefont {Baranov},\ and\ \citenamefont
  {Zoller}}]{diehl2011topology}%
  \BibitemOpen
  \bibfield  {author} {\bibinfo {author} {\bibfnamefont {S.}~\bibnamefont
  {Diehl}}, \bibinfo {author} {\bibfnamefont {E.}~\bibnamefont {Rico}},
  \bibinfo {author} {\bibfnamefont {M.~A.}\ \bibnamefont {Baranov}},\ and\
  \bibinfo {author} {\bibfnamefont {P.}~\bibnamefont {Zoller}},\ }\href
  {https://www.nature.com/articles/nphys2106} {\bibfield  {journal} {\bibinfo
  {journal} {Nature Physics}\ }\textbf {\bibinfo {volume} {7}},\ \bibinfo
  {pages} {971} (\bibinfo {year} {2011})}\BibitemShut {NoStop}%
\bibitem [{\citenamefont {Song}\ \emph {et~al.}(2019)\citenamefont {Song},
  \citenamefont {Yao},\ and\ \citenamefont {Wang}}]{PhysRevLett.123.170401}%
  \BibitemOpen
  \bibfield  {author} {\bibinfo {author} {\bibfnamefont {F.}~\bibnamefont
  {Song}}, \bibinfo {author} {\bibfnamefont {S.}~\bibnamefont {Yao}},\ and\
  \bibinfo {author} {\bibfnamefont {Z.}~\bibnamefont {Wang}},\ }\href
  {https://doi.org/10.1103/PhysRevLett.123.170401} {\bibfield  {journal}
  {\bibinfo  {journal} {Phys. Rev. Lett.}\ }\textbf {\bibinfo {volume} {123}},\
  \bibinfo {pages} {170401} (\bibinfo {year} {2019})}\BibitemShut {NoStop}%
\bibitem [{\citenamefont {Lieu}\ \emph {et~al.}(2020)\citenamefont {Lieu},
  \citenamefont {McGinley},\ and\ \citenamefont
  {Cooper}}]{PhysRevLett.124.040401}%
  \BibitemOpen
  \bibfield  {author} {\bibinfo {author} {\bibfnamefont {S.}~\bibnamefont
  {Lieu}}, \bibinfo {author} {\bibfnamefont {M.}~\bibnamefont {McGinley}},\
  and\ \bibinfo {author} {\bibfnamefont {N.~R.}\ \bibnamefont {Cooper}},\
  }\href {https://doi.org/10.1103/PhysRevLett.124.040401} {\bibfield  {journal}
  {\bibinfo  {journal} {Phys. Rev. Lett.}\ }\textbf {\bibinfo {volume} {124}},\
  \bibinfo {pages} {040401} (\bibinfo {year} {2020})}\BibitemShut {NoStop}%
\bibitem [{\citenamefont {Ma}\ \emph {et~al.}(2022)\citenamefont {Ma},
  \citenamefont {Horoshko}, \citenamefont {Yu},\ and\ \citenamefont
  {Kilin}}]{PhysRevA.105.053718}%
  \BibitemOpen
  \bibfield  {author} {\bibinfo {author} {\bibfnamefont {T.-t.}\ \bibnamefont
  {Ma}}, \bibinfo {author} {\bibfnamefont {D.~B.}\ \bibnamefont {Horoshko}},
  \bibinfo {author} {\bibfnamefont {C.-s.}\ \bibnamefont {Yu}},\ and\ \bibinfo
  {author} {\bibfnamefont {S.~Y.}\ \bibnamefont {Kilin}},\ }\href
  {https://doi.org/10.1103/PhysRevA.105.053718} {\bibfield  {journal} {\bibinfo
   {journal} {Phys. Rev. A}\ }\textbf {\bibinfo {volume} {105}},\ \bibinfo
  {pages} {053718} (\bibinfo {year} {2022})}\BibitemShut {NoStop}%
\bibitem [{\citenamefont {Kawabata}\ \emph {et~al.}(2019)\citenamefont
  {Kawabata}, \citenamefont {Shiozaki}, \citenamefont {Ueda},\ and\
  \citenamefont {Sato}}]{PhysRevX.9.041015}%
  \BibitemOpen
  \bibfield  {author} {\bibinfo {author} {\bibfnamefont {K.}~\bibnamefont
  {Kawabata}}, \bibinfo {author} {\bibfnamefont {K.}~\bibnamefont {Shiozaki}},
  \bibinfo {author} {\bibfnamefont {M.}~\bibnamefont {Ueda}},\ and\ \bibinfo
  {author} {\bibfnamefont {M.}~\bibnamefont {Sato}},\ }\href
  {https://doi.org/10.1103/PhysRevX.9.041015} {\bibfield  {journal} {\bibinfo
  {journal} {Phys. Rev. X}\ }\textbf {\bibinfo {volume} {9}},\ \bibinfo {pages}
  {041015} (\bibinfo {year} {2019})}\BibitemShut {NoStop}%
\bibitem [{\citenamefont {Gong}\ \emph {et~al.}(2018)\citenamefont {Gong},
  \citenamefont {Ashida}, \citenamefont {Kawabata}, \citenamefont {Takasan},
  \citenamefont {Higashikawa},\ and\ \citenamefont {Ueda}}]{PhysRevX.8.031079}%
  \BibitemOpen
  \bibfield  {author} {\bibinfo {author} {\bibfnamefont {Z.}~\bibnamefont
  {Gong}}, \bibinfo {author} {\bibfnamefont {Y.}~\bibnamefont {Ashida}},
  \bibinfo {author} {\bibfnamefont {K.}~\bibnamefont {Kawabata}}, \bibinfo
  {author} {\bibfnamefont {K.}~\bibnamefont {Takasan}}, \bibinfo {author}
  {\bibfnamefont {S.}~\bibnamefont {Higashikawa}},\ and\ \bibinfo {author}
  {\bibfnamefont {M.}~\bibnamefont {Ueda}},\ }\href
  {https://doi.org/10.1103/PhysRevX.8.031079} {\bibfield  {journal} {\bibinfo
  {journal} {Phys. Rev. X}\ }\textbf {\bibinfo {volume} {8}},\ \bibinfo {pages}
  {031079} (\bibinfo {year} {2018})}\BibitemShut {NoStop}%
\bibitem [{\citenamefont {Kunst}\ \emph {et~al.}(2018)\citenamefont {Kunst},
  \citenamefont {Edvardsson}, \citenamefont {Budich},\ and\ \citenamefont
  {Bergholtz}}]{PhysRevLett.121.026808}%
  \BibitemOpen
  \bibfield  {author} {\bibinfo {author} {\bibfnamefont {F.~K.}\ \bibnamefont
  {Kunst}}, \bibinfo {author} {\bibfnamefont {E.}~\bibnamefont {Edvardsson}},
  \bibinfo {author} {\bibfnamefont {J.~C.}\ \bibnamefont {Budich}},\ and\
  \bibinfo {author} {\bibfnamefont {E.~J.}\ \bibnamefont {Bergholtz}},\ }\href
  {https://doi.org/10.1103/PhysRevLett.121.026808} {\bibfield  {journal}
  {\bibinfo  {journal} {Phys. Rev. Lett.}\ }\textbf {\bibinfo {volume} {121}},\
  \bibinfo {pages} {026808} (\bibinfo {year} {2018})}\BibitemShut {NoStop}%
\bibitem [{\citenamefont {Jin}\ and\ \citenamefont
  {Song}(2019)}]{PhysRevB.99.081103}%
  \BibitemOpen
  \bibfield  {author} {\bibinfo {author} {\bibfnamefont {L.}~\bibnamefont
  {Jin}}\ and\ \bibinfo {author} {\bibfnamefont {Z.}~\bibnamefont {Song}},\
  }\href {https://doi.org/10.1103/PhysRevB.99.081103} {\bibfield  {journal}
  {\bibinfo  {journal} {Phys. Rev. B}\ }\textbf {\bibinfo {volume} {99}},\
  \bibinfo {pages} {081103(R)} (\bibinfo {year} {2019})}\BibitemShut {NoStop}%
\bibitem [{\citenamefont {Edvardsson}\ \emph {et~al.}(2019)\citenamefont
  {Edvardsson}, \citenamefont {Kunst},\ and\ \citenamefont
  {Bergholtz}}]{PhysRevB.99.081302}%
  \BibitemOpen
  \bibfield  {author} {\bibinfo {author} {\bibfnamefont {E.}~\bibnamefont
  {Edvardsson}}, \bibinfo {author} {\bibfnamefont {F.~K.}\ \bibnamefont
  {Kunst}},\ and\ \bibinfo {author} {\bibfnamefont {E.~J.}\ \bibnamefont
  {Bergholtz}},\ }\href {https://doi.org/10.1103/PhysRevB.99.081302} {\bibfield
   {journal} {\bibinfo  {journal} {Phys. Rev. B}\ }\textbf {\bibinfo {volume}
  {99}},\ \bibinfo {pages} {081302(R)} (\bibinfo {year} {2019})}\BibitemShut
  {NoStop}%
\bibitem [{\citenamefont {Imura}\ and\ \citenamefont
  {Takane}(2019)}]{PhysRevB.100.165430}%
  \BibitemOpen
  \bibfield  {author} {\bibinfo {author} {\bibfnamefont {K.-I.}\ \bibnamefont
  {Imura}}\ and\ \bibinfo {author} {\bibfnamefont {Y.}~\bibnamefont {Takane}},\
  }\href {https://doi.org/10.1103/PhysRevB.100.165430} {\bibfield  {journal}
  {\bibinfo  {journal} {Phys. Rev. B}\ }\textbf {\bibinfo {volume} {100}},\
  \bibinfo {pages} {165430} (\bibinfo {year} {2019})}\BibitemShut {NoStop}%
\bibitem [{\citenamefont {Herviou}\ \emph {et~al.}(2019)\citenamefont
  {Herviou}, \citenamefont {Bardarson},\ and\ \citenamefont
  {Regnault}}]{PhysRevA.99.052118}%
  \BibitemOpen
  \bibfield  {author} {\bibinfo {author} {\bibfnamefont {L.}~\bibnamefont
  {Herviou}}, \bibinfo {author} {\bibfnamefont {J.~H.}\ \bibnamefont
  {Bardarson}},\ and\ \bibinfo {author} {\bibfnamefont {N.}~\bibnamefont
  {Regnault}},\ }\href {https://doi.org/10.1103/PhysRevA.99.052118} {\bibfield
  {journal} {\bibinfo  {journal} {Phys. Rev. A}\ }\textbf {\bibinfo {volume}
  {99}},\ \bibinfo {pages} {052118} (\bibinfo {year} {2019})}\BibitemShut
  {NoStop}%
\bibitem [{\citenamefont {Xiao}\ \emph
  {et~al.}(2020{\natexlab{a}})\citenamefont {Xiao}, \citenamefont {Deng},
  \citenamefont {Wang}, \citenamefont {Zhu}, \citenamefont {Wang},
  \citenamefont {Yi},\ and\ \citenamefont {Xue}}]{xiao2020non}%
  \BibitemOpen
  \bibfield  {author} {\bibinfo {author} {\bibfnamefont {L.}~\bibnamefont
  {Xiao}}, \bibinfo {author} {\bibfnamefont {T.}~\bibnamefont {Deng}}, \bibinfo
  {author} {\bibfnamefont {K.}~\bibnamefont {Wang}}, \bibinfo {author}
  {\bibfnamefont {G.}~\bibnamefont {Zhu}}, \bibinfo {author} {\bibfnamefont
  {Z.}~\bibnamefont {Wang}}, \bibinfo {author} {\bibfnamefont {W.}~\bibnamefont
  {Yi}},\ and\ \bibinfo {author} {\bibfnamefont {P.}~\bibnamefont {Xue}},\
  }\href {https://www.nature.com/articles/s41567-020-0836-6} {\bibfield
  {journal} {\bibinfo  {journal} {Nature Physics}\ }\textbf {\bibinfo {volume}
  {16}},\ \bibinfo {pages} {761} (\bibinfo {year}
  {2020}{\natexlab{a}})}\BibitemShut {NoStop}%
\bibitem [{\citenamefont {Zhu}\ \emph {et~al.}(2020)\citenamefont {Zhu},
  \citenamefont {Wang}, \citenamefont {Gupta}, \citenamefont {Zhang},
  \citenamefont {Xie}, \citenamefont {Lu},\ and\ \citenamefont
  {Chen}}]{PhysRevResearch.2.013280}%
  \BibitemOpen
  \bibfield  {author} {\bibinfo {author} {\bibfnamefont {X.}~\bibnamefont
  {Zhu}}, \bibinfo {author} {\bibfnamefont {H.}~\bibnamefont {Wang}}, \bibinfo
  {author} {\bibfnamefont {S.~K.}\ \bibnamefont {Gupta}}, \bibinfo {author}
  {\bibfnamefont {H.}~\bibnamefont {Zhang}}, \bibinfo {author} {\bibfnamefont
  {B.}~\bibnamefont {Xie}}, \bibinfo {author} {\bibfnamefont {M.}~\bibnamefont
  {Lu}},\ and\ \bibinfo {author} {\bibfnamefont {Y.}~\bibnamefont {Chen}},\
  }\href {https://doi.org/10.1103/PhysRevResearch.2.013280} {\bibfield
  {journal} {\bibinfo  {journal} {Phys. Rev. Res.}\ }\textbf {\bibinfo {volume}
  {2}},\ \bibinfo {pages} {013280} (\bibinfo {year} {2020})}\BibitemShut
  {NoStop}%
\bibitem [{\citenamefont {Yang}\ \emph {et~al.}(2020)\citenamefont {Yang},
  \citenamefont {Zhang}, \citenamefont {Fang},\ and\ \citenamefont
  {Hu}}]{PhysRevLett.125.226402}%
  \BibitemOpen
  \bibfield  {author} {\bibinfo {author} {\bibfnamefont {Z.}~\bibnamefont
  {Yang}}, \bibinfo {author} {\bibfnamefont {K.}~\bibnamefont {Zhang}},
  \bibinfo {author} {\bibfnamefont {C.}~\bibnamefont {Fang}},\ and\ \bibinfo
  {author} {\bibfnamefont {J.}~\bibnamefont {Hu}},\ }\href
  {https://doi.org/10.1103/PhysRevLett.125.226402} {\bibfield  {journal}
  {\bibinfo  {journal} {Phys. Rev. Lett.}\ }\textbf {\bibinfo {volume} {125}},\
  \bibinfo {pages} {226402} (\bibinfo {year} {2020})}\BibitemShut {NoStop}%
\bibitem [{\citenamefont {Zirnstein}\ \emph {et~al.}(2021)\citenamefont
  {Zirnstein}, \citenamefont {Refael},\ and\ \citenamefont
  {Rosenow}}]{PhysRevLett.126.216407}%
  \BibitemOpen
  \bibfield  {author} {\bibinfo {author} {\bibfnamefont {H.-G.}\ \bibnamefont
  {Zirnstein}}, \bibinfo {author} {\bibfnamefont {G.}~\bibnamefont {Refael}},\
  and\ \bibinfo {author} {\bibfnamefont {B.}~\bibnamefont {Rosenow}},\ }\href
  {https://doi.org/10.1103/PhysRevLett.126.216407} {\bibfield  {journal}
  {\bibinfo  {journal} {Phys. Rev. Lett.}\ }\textbf {\bibinfo {volume} {126}},\
  \bibinfo {pages} {216407} (\bibinfo {year} {2021})}\BibitemShut {NoStop}%
\bibitem [{\citenamefont {Cao}\ \emph {et~al.}(2021)\citenamefont {Cao},
  \citenamefont {Li},\ and\ \citenamefont {Yang}}]{PhysRevB.103.075126}%
  \BibitemOpen
  \bibfield  {author} {\bibinfo {author} {\bibfnamefont {Y.}~\bibnamefont
  {Cao}}, \bibinfo {author} {\bibfnamefont {Y.}~\bibnamefont {Li}},\ and\
  \bibinfo {author} {\bibfnamefont {X.}~\bibnamefont {Yang}},\ }\href
  {https://doi.org/10.1103/PhysRevB.103.075126} {\bibfield  {journal} {\bibinfo
   {journal} {Phys. Rev. B}\ }\textbf {\bibinfo {volume} {103}},\ \bibinfo
  {pages} {075126} (\bibinfo {year} {2021})}\BibitemShut {NoStop}%
\bibitem [{\citenamefont {Hwang}\ and\ \citenamefont
  {Obuse}(2023)}]{PhysRevB.108.L121302}%
  \BibitemOpen
  \bibfield  {author} {\bibinfo {author} {\bibfnamefont {G.}~\bibnamefont
  {Hwang}}\ and\ \bibinfo {author} {\bibfnamefont {H.}~\bibnamefont {Obuse}},\
  }\href {https://doi.org/10.1103/PhysRevB.108.L121302} {\bibfield  {journal}
  {\bibinfo  {journal} {Phys. Rev. B}\ }\textbf {\bibinfo {volume} {108}},\
  \bibinfo {pages} {L121302} (\bibinfo {year} {2023})}\BibitemShut {NoStop}%
\bibitem [{\citenamefont {Jiang}\ \emph {et~al.}(2019)\citenamefont {Jiang},
  \citenamefont {Lang}, \citenamefont {Yang}, \citenamefont {Zhu},\ and\
  \citenamefont {Chen}}]{PhysRevB.100.054301}%
  \BibitemOpen
  \bibfield  {author} {\bibinfo {author} {\bibfnamefont {H.}~\bibnamefont
  {Jiang}}, \bibinfo {author} {\bibfnamefont {L.-J.}\ \bibnamefont {Lang}},
  \bibinfo {author} {\bibfnamefont {C.}~\bibnamefont {Yang}}, \bibinfo {author}
  {\bibfnamefont {S.-L.}\ \bibnamefont {Zhu}},\ and\ \bibinfo {author}
  {\bibfnamefont {S.}~\bibnamefont {Chen}},\ }\href
  {https://doi.org/10.1103/PhysRevB.100.054301} {\bibfield  {journal} {\bibinfo
   {journal} {Phys. Rev. B}\ }\textbf {\bibinfo {volume} {100}},\ \bibinfo
  {pages} {054301} (\bibinfo {year} {2019})}\BibitemShut {NoStop}%
\bibitem [{\citenamefont {Borgnia}\ \emph {et~al.}(2020)\citenamefont
  {Borgnia}, \citenamefont {Kruchkov},\ and\ \citenamefont
  {Slager}}]{PhysRevLett.124.056802}%
  \BibitemOpen
  \bibfield  {author} {\bibinfo {author} {\bibfnamefont {D.~S.}\ \bibnamefont
  {Borgnia}}, \bibinfo {author} {\bibfnamefont {A.~J.}\ \bibnamefont
  {Kruchkov}},\ and\ \bibinfo {author} {\bibfnamefont {R.-J.}\ \bibnamefont
  {Slager}},\ }\href {https://doi.org/10.1103/PhysRevLett.124.056802}
  {\bibfield  {journal} {\bibinfo  {journal} {Phys. Rev. Lett.}\ }\textbf
  {\bibinfo {volume} {124}},\ \bibinfo {pages} {056802} (\bibinfo {year}
  {2020})}\BibitemShut {NoStop}%
\bibitem [{\citenamefont {Okuma}\ \emph {et~al.}(2020)\citenamefont {Okuma},
  \citenamefont {Kawabata}, \citenamefont {Shiozaki},\ and\ \citenamefont
  {Sato}}]{Okuma2020}%
  \BibitemOpen
  \bibfield  {author} {\bibinfo {author} {\bibfnamefont {N.}~\bibnamefont
  {Okuma}}, \bibinfo {author} {\bibfnamefont {K.}~\bibnamefont {Kawabata}},
  \bibinfo {author} {\bibfnamefont {K.}~\bibnamefont {Shiozaki}},\ and\
  \bibinfo {author} {\bibfnamefont {M.}~\bibnamefont {Sato}},\ }\href
  {https://doi.org/10.1103/PhysRevLett.124.086801} {\bibfield  {journal}
  {\bibinfo  {journal} {Phys. Rev. Lett.}\ }\textbf {\bibinfo {volume} {124}},\
  \bibinfo {pages} {086801} (\bibinfo {year} {2020})}\BibitemShut {NoStop}%
\bibitem [{\citenamefont {Zhang}\ \emph {et~al.}(2020)\citenamefont {Zhang},
  \citenamefont {Yang},\ and\ \citenamefont {Fang}}]{PhysRevLett.125.126402}%
  \BibitemOpen
  \bibfield  {author} {\bibinfo {author} {\bibfnamefont {K.}~\bibnamefont
  {Zhang}}, \bibinfo {author} {\bibfnamefont {Z.}~\bibnamefont {Yang}},\ and\
  \bibinfo {author} {\bibfnamefont {C.}~\bibnamefont {Fang}},\ }\href
  {https://doi.org/10.1103/PhysRevLett.125.126402} {\bibfield  {journal}
  {\bibinfo  {journal} {Phys. Rev. Lett.}\ }\textbf {\bibinfo {volume} {125}},\
  \bibinfo {pages} {126402} (\bibinfo {year} {2020})}\BibitemShut {NoStop}%
\bibitem [{\citenamefont {Li}\ \emph {et~al.}(2020)\citenamefont {Li},
  \citenamefont {Lee}, \citenamefont {Mu},\ and\ \citenamefont
  {Gong}}]{li2020critical}%
  \BibitemOpen
  \bibfield  {author} {\bibinfo {author} {\bibfnamefont {L.}~\bibnamefont
  {Li}}, \bibinfo {author} {\bibfnamefont {C.~H.}\ \bibnamefont {Lee}},
  \bibinfo {author} {\bibfnamefont {S.}~\bibnamefont {Mu}},\ and\ \bibinfo
  {author} {\bibfnamefont {J.}~\bibnamefont {Gong}},\ }\href
  {https://www.nature.com/articles/s41467-020-18917-4} {\bibfield  {journal}
  {\bibinfo  {journal} {Nature communications}\ }\textbf {\bibinfo {volume}
  {11}},\ \bibinfo {pages} {5491} (\bibinfo {year} {2020})}\BibitemShut
  {NoStop}%
\bibitem [{\citenamefont {Longhi}(2020)}]{PhysRevB.102.201103}%
  \BibitemOpen
  \bibfield  {author} {\bibinfo {author} {\bibfnamefont {S.}~\bibnamefont
  {Longhi}},\ }\href {https://doi.org/10.1103/PhysRevB.102.201103} {\bibfield
  {journal} {\bibinfo  {journal} {Phys. Rev. B}\ }\textbf {\bibinfo {volume}
  {102}},\ \bibinfo {pages} {201103(R)} (\bibinfo {year} {2020})}\BibitemShut
  {NoStop}%
\bibitem [{\citenamefont {Okugawa}\ \emph {et~al.}(2020)\citenamefont
  {Okugawa}, \citenamefont {Takahashi},\ and\ \citenamefont
  {Yokomizo}}]{PhysRevB.102.241202}%
  \BibitemOpen
  \bibfield  {author} {\bibinfo {author} {\bibfnamefont {R.}~\bibnamefont
  {Okugawa}}, \bibinfo {author} {\bibfnamefont {R.}~\bibnamefont {Takahashi}},\
  and\ \bibinfo {author} {\bibfnamefont {K.}~\bibnamefont {Yokomizo}},\ }\href
  {https://doi.org/10.1103/PhysRevB.102.241202} {\bibfield  {journal} {\bibinfo
   {journal} {Phys. Rev. B}\ }\textbf {\bibinfo {volume} {102}},\ \bibinfo
  {pages} {241202(R)} (\bibinfo {year} {2020})}\BibitemShut {NoStop}%
\bibitem [{\citenamefont {Zhang}\ \emph {et~al.}(2021)\citenamefont {Zhang},
  \citenamefont {Tian}, \citenamefont {Jiang}, \citenamefont {Lu},\ and\
  \citenamefont {Chen}}]{zhang2021observation}%
  \BibitemOpen
  \bibfield  {author} {\bibinfo {author} {\bibfnamefont {X.}~\bibnamefont
  {Zhang}}, \bibinfo {author} {\bibfnamefont {Y.}~\bibnamefont {Tian}},
  \bibinfo {author} {\bibfnamefont {J.-H.}\ \bibnamefont {Jiang}}, \bibinfo
  {author} {\bibfnamefont {M.-H.}\ \bibnamefont {Lu}},\ and\ \bibinfo {author}
  {\bibfnamefont {Y.-F.}\ \bibnamefont {Chen}},\ }\href
  {https://www.nature.com/articles/s41467-021-25716-y} {\bibfield  {journal}
  {\bibinfo  {journal} {Nature communications}\ }\textbf {\bibinfo {volume}
  {12}},\ \bibinfo {pages} {5377} (\bibinfo {year} {2021})}\BibitemShut
  {NoStop}%
\bibitem [{\citenamefont {Yokomizo}\ and\ \citenamefont
  {Murakami}(2021)}]{PhysRevB.104.165117}%
  \BibitemOpen
  \bibfield  {author} {\bibinfo {author} {\bibfnamefont {K.}~\bibnamefont
  {Yokomizo}}\ and\ \bibinfo {author} {\bibfnamefont {S.}~\bibnamefont
  {Murakami}},\ }\href {https://doi.org/10.1103/PhysRevB.104.165117} {\bibfield
   {journal} {\bibinfo  {journal} {Phys. Rev. B}\ }\textbf {\bibinfo {volume}
  {104}},\ \bibinfo {pages} {165117} (\bibinfo {year} {2021})}\BibitemShut
  {NoStop}%
\bibitem [{\citenamefont {Longhi}(2021)}]{PhysRevB.104.125109}%
  \BibitemOpen
  \bibfield  {author} {\bibinfo {author} {\bibfnamefont {S.}~\bibnamefont
  {Longhi}},\ }\href {https://doi.org/10.1103/PhysRevB.104.125109} {\bibfield
  {journal} {\bibinfo  {journal} {Phys. Rev. B}\ }\textbf {\bibinfo {volume}
  {104}},\ \bibinfo {pages} {125109} (\bibinfo {year} {2021})}\BibitemShut
  {NoStop}%
\bibitem [{\citenamefont {Claes}\ and\ \citenamefont
  {Hughes}(2021)}]{PhysRevB.103.L140201}%
  \BibitemOpen
  \bibfield  {author} {\bibinfo {author} {\bibfnamefont {J.}~\bibnamefont
  {Claes}}\ and\ \bibinfo {author} {\bibfnamefont {T.~L.}\ \bibnamefont
  {Hughes}},\ }\href {https://doi.org/10.1103/PhysRevB.103.L140201} {\bibfield
  {journal} {\bibinfo  {journal} {Phys. Rev. B}\ }\textbf {\bibinfo {volume}
  {103}},\ \bibinfo {pages} {L140201} (\bibinfo {year} {2021})}\BibitemShut
  {NoStop}%
\bibitem [{\citenamefont {Okuma}\ and\ \citenamefont
  {Sato}(2021)}]{PhysRevB.103.085428}%
  \BibitemOpen
  \bibfield  {author} {\bibinfo {author} {\bibfnamefont {N.}~\bibnamefont
  {Okuma}}\ and\ \bibinfo {author} {\bibfnamefont {M.}~\bibnamefont {Sato}},\
  }\href {https://doi.org/10.1103/PhysRevB.103.085428} {\bibfield  {journal}
  {\bibinfo  {journal} {Phys. Rev. B}\ }\textbf {\bibinfo {volume} {103}},\
  \bibinfo {pages} {085428} (\bibinfo {year} {2021})}\BibitemShut {NoStop}%
\bibitem [{\citenamefont {Zhang}\ \emph {et~al.}(2022)\citenamefont {Zhang},
  \citenamefont {Yang},\ and\ \citenamefont {Fang}}]{zhang2022universal}%
  \BibitemOpen
  \bibfield  {author} {\bibinfo {author} {\bibfnamefont {K.}~\bibnamefont
  {Zhang}}, \bibinfo {author} {\bibfnamefont {Z.}~\bibnamefont {Yang}},\ and\
  \bibinfo {author} {\bibfnamefont {C.}~\bibnamefont {Fang}},\ }\href
  {https://www.nature.com/articles/s41467-022-30161-6} {\bibfield  {journal}
  {\bibinfo  {journal} {Nature communications}\ }\textbf {\bibinfo {volume}
  {13}},\ \bibinfo {pages} {2496} (\bibinfo {year} {2022})}\BibitemShut
  {NoStop}%
\bibitem [{\citenamefont {Zeng}(2022)}]{PhysRevB.106.235411}%
  \BibitemOpen
  \bibfield  {author} {\bibinfo {author} {\bibfnamefont {Q.-B.}\ \bibnamefont
  {Zeng}},\ }\href {https://doi.org/10.1103/PhysRevB.106.235411} {\bibfield
  {journal} {\bibinfo  {journal} {Phys. Rev. B}\ }\textbf {\bibinfo {volume}
  {106}},\ \bibinfo {pages} {235411} (\bibinfo {year} {2022})}\BibitemShut
  {NoStop}%
\bibitem [{\citenamefont {Roccati}\ \emph {et~al.}(2024)\citenamefont
  {Roccati}, \citenamefont {Bello}, \citenamefont {Gong}, \citenamefont {Ueda},
  \citenamefont {Ciccarello}, \citenamefont {Chenu},\ and\ \citenamefont
  {Carollo}}]{roccati2024hermitian}%
  \BibitemOpen
  \bibfield  {author} {\bibinfo {author} {\bibfnamefont {F.}~\bibnamefont
  {Roccati}}, \bibinfo {author} {\bibfnamefont {M.}~\bibnamefont {Bello}},
  \bibinfo {author} {\bibfnamefont {Z.}~\bibnamefont {Gong}}, \bibinfo {author}
  {\bibfnamefont {M.}~\bibnamefont {Ueda}}, \bibinfo {author} {\bibfnamefont
  {F.}~\bibnamefont {Ciccarello}}, \bibinfo {author} {\bibfnamefont
  {A.}~\bibnamefont {Chenu}},\ and\ \bibinfo {author} {\bibfnamefont
  {A.}~\bibnamefont {Carollo}},\ }\href
  {https://www.nature.com/articles/s41467-024-46471-w} {\bibfield  {journal}
  {\bibinfo  {journal} {Nature Communications}\ }\textbf {\bibinfo {volume}
  {15}},\ \bibinfo {pages} {2400} (\bibinfo {year} {2024})}\BibitemShut
  {NoStop}%
\bibitem [{\citenamefont {Nakai}\ \emph {et~al.}(2024)\citenamefont {Nakai},
  \citenamefont {Okuma}, \citenamefont {Nakamura}, \citenamefont {Shimomura},\
  and\ \citenamefont {Sato}}]{Nakai2024}%
  \BibitemOpen
  \bibfield  {author} {\bibinfo {author} {\bibfnamefont {Y.~O.}\ \bibnamefont
  {Nakai}}, \bibinfo {author} {\bibfnamefont {N.}~\bibnamefont {Okuma}},
  \bibinfo {author} {\bibfnamefont {D.}~\bibnamefont {Nakamura}}, \bibinfo
  {author} {\bibfnamefont {K.}~\bibnamefont {Shimomura}},\ and\ \bibinfo
  {author} {\bibfnamefont {M.}~\bibnamefont {Sato}},\ }\href
  {https://doi.org/10.1103/PhysRevB.109.144203} {\bibfield  {journal} {\bibinfo
   {journal} {Phys. Rev. B}\ }\textbf {\bibinfo {volume} {109}},\ \bibinfo
  {pages} {144203} (\bibinfo {year} {2024})}\BibitemShut {NoStop}%
\bibitem [{\citenamefont {Hatano}\ and\ \citenamefont
  {Nelson}(1996)}]{PhysRevLett.77.570}%
  \BibitemOpen
  \bibfield  {author} {\bibinfo {author} {\bibfnamefont {N.}~\bibnamefont
  {Hatano}}\ and\ \bibinfo {author} {\bibfnamefont {D.~R.}\ \bibnamefont
  {Nelson}},\ }\href {https://doi.org/10.1103/PhysRevLett.77.570} {\bibfield
  {journal} {\bibinfo  {journal} {Phys. Rev. Lett.}\ }\textbf {\bibinfo
  {volume} {77}},\ \bibinfo {pages} {570} (\bibinfo {year} {1996})}\BibitemShut
  {NoStop}%
\bibitem [{\citenamefont {Hatano}\ and\ \citenamefont
  {Nelson}(1997)}]{PhysRevB.56.8651}%
  \BibitemOpen
  \bibfield  {author} {\bibinfo {author} {\bibfnamefont {N.}~\bibnamefont
  {Hatano}}\ and\ \bibinfo {author} {\bibfnamefont {D.~R.}\ \bibnamefont
  {Nelson}},\ }\href {https://doi.org/10.1103/PhysRevB.56.8651} {\bibfield
  {journal} {\bibinfo  {journal} {Phys. Rev. B}\ }\textbf {\bibinfo {volume}
  {56}},\ \bibinfo {pages} {8651} (\bibinfo {year} {1997})}\BibitemShut
  {NoStop}%
\bibitem [{\citenamefont {Hatano}\ and\ \citenamefont
  {Nelson}(1998)}]{PhysRevB.58.8384}%
  \BibitemOpen
  \bibfield  {author} {\bibinfo {author} {\bibfnamefont {N.}~\bibnamefont
  {Hatano}}\ and\ \bibinfo {author} {\bibfnamefont {D.~R.}\ \bibnamefont
  {Nelson}},\ }\href {https://doi.org/10.1103/PhysRevB.58.8384} {\bibfield
  {journal} {\bibinfo  {journal} {Phys. Rev. B}\ }\textbf {\bibinfo {volume}
  {58}},\ \bibinfo {pages} {8384} (\bibinfo {year} {1998})}\BibitemShut
  {NoStop}%
\bibitem [{\citenamefont {Mochizuki}\ \emph {et~al.}(2016)\citenamefont
  {Mochizuki}, \citenamefont {Kim},\ and\ \citenamefont
  {Obuse}}]{PhysRevA.93.062116}%
  \BibitemOpen
  \bibfield  {author} {\bibinfo {author} {\bibfnamefont {K.}~\bibnamefont
  {Mochizuki}}, \bibinfo {author} {\bibfnamefont {D.}~\bibnamefont {Kim}},\
  and\ \bibinfo {author} {\bibfnamefont {H.}~\bibnamefont {Obuse}},\ }\href
  {https://doi.org/10.1103/PhysRevA.93.062116} {\bibfield  {journal} {\bibinfo
  {journal} {Phys. Rev. A}\ }\textbf {\bibinfo {volume} {93}},\ \bibinfo
  {pages} {062116} (\bibinfo {year} {2016})}\BibitemShut {NoStop}%
\bibitem [{\citenamefont {Xiao}\ \emph {et~al.}(2017)\citenamefont {Xiao},
  \citenamefont {Zhan}, \citenamefont {Bian}, \citenamefont {Wang},
  \citenamefont {Zhang}, \citenamefont {Wang}, \citenamefont {Li},
  \citenamefont {Mochizuki}, \citenamefont {Kim}, \citenamefont {Kawakami}
  \emph {et~al.}}]{xiao2017observation}%
  \BibitemOpen
  \bibfield  {author} {\bibinfo {author} {\bibfnamefont {L.}~\bibnamefont
  {Xiao}}, \bibinfo {author} {\bibfnamefont {X.}~\bibnamefont {Zhan}}, \bibinfo
  {author} {\bibfnamefont {Z.}~\bibnamefont {Bian}}, \bibinfo {author}
  {\bibfnamefont {K.}~\bibnamefont {Wang}}, \bibinfo {author} {\bibfnamefont
  {X.}~\bibnamefont {Zhang}}, \bibinfo {author} {\bibfnamefont
  {X.}~\bibnamefont {Wang}}, \bibinfo {author} {\bibfnamefont {J.}~\bibnamefont
  {Li}}, \bibinfo {author} {\bibfnamefont {K.}~\bibnamefont {Mochizuki}},
  \bibinfo {author} {\bibfnamefont {D.}~\bibnamefont {Kim}}, \bibinfo {author}
  {\bibfnamefont {N.}~\bibnamefont {Kawakami}}, \emph {et~al.},\ }\href
  {https://www.nature.com/articles/nphys4204} {\bibfield  {journal} {\bibinfo
  {journal} {Nature Physics}\ }\textbf {\bibinfo {volume} {13}},\ \bibinfo
  {pages} {1117} (\bibinfo {year} {2017})}\BibitemShut {NoStop}%
\bibitem [{\citenamefont {Kawasaki}\ \emph {et~al.}(2020)\citenamefont
  {Kawasaki}, \citenamefont {Mochizuki}, \citenamefont {Kawakami},\ and\
  \citenamefont {Obuse}}]{kawasaki2020bulk}%
  \BibitemOpen
  \bibfield  {author} {\bibinfo {author} {\bibfnamefont {M.}~\bibnamefont
  {Kawasaki}}, \bibinfo {author} {\bibfnamefont {K.}~\bibnamefont {Mochizuki}},
  \bibinfo {author} {\bibfnamefont {N.}~\bibnamefont {Kawakami}},\ and\
  \bibinfo {author} {\bibfnamefont {H.}~\bibnamefont {Obuse}},\ }\href
  {https://academic.oup.com/ptep/article/2020/12/12A105/5875997?login=true}
  {\bibfield  {journal} {\bibinfo  {journal} {Progress of Theoretical and
  Experimental Physics}\ }\textbf {\bibinfo {volume} {2020}},\ \bibinfo {pages}
  {12A105} (\bibinfo {year} {2020})}\BibitemShut {NoStop}%
\bibitem [{\citenamefont {Mochizuki}\ \emph {et~al.}(2020)\citenamefont
  {Mochizuki}, \citenamefont {Kim}, \citenamefont {Kawakami},\ and\
  \citenamefont {Obuse}}]{Mochizuki2020}%
  \BibitemOpen
  \bibfield  {author} {\bibinfo {author} {\bibfnamefont {K.}~\bibnamefont
  {Mochizuki}}, \bibinfo {author} {\bibfnamefont {D.}~\bibnamefont {Kim}},
  \bibinfo {author} {\bibfnamefont {N.}~\bibnamefont {Kawakami}},\ and\
  \bibinfo {author} {\bibfnamefont {H.}~\bibnamefont {Obuse}},\ }\href
  {https://doi.org/10.1103/PhysRevA.102.062202} {\bibfield  {journal} {\bibinfo
   {journal} {Phys. Rev. A}\ }\textbf {\bibinfo {volume} {102}},\ \bibinfo
  {pages} {062202} (\bibinfo {year} {2020})}\BibitemShut {NoStop}%
\bibitem [{\citenamefont {Xiao}\ \emph
  {et~al.}(2020{\natexlab{b}})\citenamefont {Xiao}, \citenamefont {Deng},
  \citenamefont {Wang}, \citenamefont {Zhu}, \citenamefont {Wang},
  \citenamefont {Yi},\ and\ \citenamefont {Xue}}]{Xiao2020}%
  \BibitemOpen
  \bibfield  {author} {\bibinfo {author} {\bibfnamefont {L.}~\bibnamefont
  {Xiao}}, \bibinfo {author} {\bibfnamefont {T.}~\bibnamefont {Deng}}, \bibinfo
  {author} {\bibfnamefont {K.}~\bibnamefont {Wang}}, \bibinfo {author}
  {\bibfnamefont {G.}~\bibnamefont {Zhu}}, \bibinfo {author} {\bibfnamefont
  {Z.}~\bibnamefont {Wang}}, \bibinfo {author} {\bibfnamefont {W.}~\bibnamefont
  {Yi}},\ and\ \bibinfo {author} {\bibfnamefont {P.}~\bibnamefont {Xue}},\
  }\href {https://doi.org/10.1038/s41567-020-0836-6} {\bibfield  {journal}
  {\bibinfo  {journal} {Nature Physics}\ }\textbf {\bibinfo {volume} {16}},\
  \bibinfo {pages} {761} (\bibinfo {year} {2020}{\natexlab{b}})}\BibitemShut
  {NoStop}%
\bibitem [{\citenamefont {Yamagishi}\ \emph
  {et~al.}(2023{\natexlab{a}})\citenamefont {Yamagishi}, \citenamefont
  {Hatano},\ and\ \citenamefont {Obuse}}]{active-quantum-particle}%
  \BibitemOpen
  \bibfield  {author} {\bibinfo {author} {\bibfnamefont {M.}~\bibnamefont
  {Yamagishi}}, \bibinfo {author} {\bibfnamefont {N.}~\bibnamefont {Hatano}},\
  and\ \bibinfo {author} {\bibfnamefont {H.}~\bibnamefont {Obuse}},\ }\href
  {https://arxiv.org/abs/2305.15319} {\bibfield  {journal} {\bibinfo  {journal}
  {arXiv preprint arXiv:2305.15319}\ } (\bibinfo {year}
  {2023}{\natexlab{a}})}\BibitemShut {NoStop}%
\bibitem [{\citenamefont {Aharonov}\ \emph {et~al.}(1993)\citenamefont
  {Aharonov}, \citenamefont {Davidovich},\ and\ \citenamefont
  {Zagury}}]{PhysRevA.48.1687}%
  \BibitemOpen
  \bibfield  {author} {\bibinfo {author} {\bibfnamefont {Y.}~\bibnamefont
  {Aharonov}}, \bibinfo {author} {\bibfnamefont {L.}~\bibnamefont
  {Davidovich}},\ and\ \bibinfo {author} {\bibfnamefont {N.}~\bibnamefont
  {Zagury}},\ }\href {https://doi.org/10.1103/PhysRevA.48.1687} {\bibfield
  {journal} {\bibinfo  {journal} {Phys. Rev. A}\ }\textbf {\bibinfo {volume}
  {48}},\ \bibinfo {pages} {1687} (\bibinfo {year} {1993})}\BibitemShut
  {NoStop}%
\bibitem [{\citenamefont {Tregenna}\ \emph {et~al.}(2003)\citenamefont
  {Tregenna}, \citenamefont {Flanagan}, \citenamefont {Maile},\ and\
  \citenamefont {Kendon}}]{tregenna2003controlling}%
  \BibitemOpen
  \bibfield  {author} {\bibinfo {author} {\bibfnamefont {B.}~\bibnamefont
  {Tregenna}}, \bibinfo {author} {\bibfnamefont {W.}~\bibnamefont {Flanagan}},
  \bibinfo {author} {\bibfnamefont {R.}~\bibnamefont {Maile}},\ and\ \bibinfo
  {author} {\bibfnamefont {V.}~\bibnamefont {Kendon}},\ }\href
  {https://iopscience.iop.org/article/10.1088/1367-2630/5/1/383/meta}
  {\bibfield  {journal} {\bibinfo  {journal} {New Journal of Physics}\ }\textbf
  {\bibinfo {volume} {5}},\ \bibinfo {pages} {83} (\bibinfo {year}
  {2003})}\BibitemShut {NoStop}%
\bibitem [{\citenamefont {Chandrashekar}\ \emph {et~al.}(2008)\citenamefont
  {Chandrashekar}, \citenamefont {Srikanth},\ and\ \citenamefont
  {Laflamme}}]{PhysRevA.77.032326}%
  \BibitemOpen
  \bibfield  {author} {\bibinfo {author} {\bibfnamefont {C.~M.}\ \bibnamefont
  {Chandrashekar}}, \bibinfo {author} {\bibfnamefont {R.}~\bibnamefont
  {Srikanth}},\ and\ \bibinfo {author} {\bibfnamefont {R.}~\bibnamefont
  {Laflamme}},\ }\href {https://doi.org/10.1103/PhysRevA.77.032326} {\bibfield
  {journal} {\bibinfo  {journal} {Phys. Rev. A}\ }\textbf {\bibinfo {volume}
  {77}},\ \bibinfo {pages} {032326} (\bibinfo {year} {2008})}\BibitemShut
  {NoStop}%
\bibitem [{\citenamefont {Lovett}\ \emph {et~al.}(2010)\citenamefont {Lovett},
  \citenamefont {Cooper}, \citenamefont {Everitt}, \citenamefont {Trevers},\
  and\ \citenamefont {Kendon}}]{PhysRevA.81.042330}%
  \BibitemOpen
  \bibfield  {author} {\bibinfo {author} {\bibfnamefont {N.~B.}\ \bibnamefont
  {Lovett}}, \bibinfo {author} {\bibfnamefont {S.}~\bibnamefont {Cooper}},
  \bibinfo {author} {\bibfnamefont {M.}~\bibnamefont {Everitt}}, \bibinfo
  {author} {\bibfnamefont {M.}~\bibnamefont {Trevers}},\ and\ \bibinfo {author}
  {\bibfnamefont {V.}~\bibnamefont {Kendon}},\ }\href
  {https://doi.org/10.1103/PhysRevA.81.042330} {\bibfield  {journal} {\bibinfo
  {journal} {Phys. Rev. A}\ }\textbf {\bibinfo {volume} {81}},\ \bibinfo
  {pages} {042330} (\bibinfo {year} {2010})}\BibitemShut {NoStop}%
\bibitem [{\citenamefont {Schreiber}\ \emph {et~al.}(2010)\citenamefont
  {Schreiber}, \citenamefont {Cassemiro}, \citenamefont
  {Poto\ifmmode~\check{c}\else \v{c}\fi{}ek}, \citenamefont {G\'abris},
  \citenamefont {Mosley}, \citenamefont {Andersson}, \citenamefont {Jex},\ and\
  \citenamefont {Silberhorn}}]{PhysRevLett.104.050502}%
  \BibitemOpen
  \bibfield  {author} {\bibinfo {author} {\bibfnamefont {A.}~\bibnamefont
  {Schreiber}}, \bibinfo {author} {\bibfnamefont {K.~N.}\ \bibnamefont
  {Cassemiro}}, \bibinfo {author} {\bibfnamefont {V.}~\bibnamefont
  {Poto\ifmmode~\check{c}\else \v{c}\fi{}ek}}, \bibinfo {author} {\bibfnamefont
  {A.}~\bibnamefont {G\'abris}}, \bibinfo {author} {\bibfnamefont {P.~J.}\
  \bibnamefont {Mosley}}, \bibinfo {author} {\bibfnamefont {E.}~\bibnamefont
  {Andersson}}, \bibinfo {author} {\bibfnamefont {I.}~\bibnamefont {Jex}},\
  and\ \bibinfo {author} {\bibfnamefont {C.}~\bibnamefont {Silberhorn}},\
  }\href {https://doi.org/10.1103/PhysRevLett.104.050502} {\bibfield  {journal}
  {\bibinfo  {journal} {Phys. Rev. Lett.}\ }\textbf {\bibinfo {volume} {104}},\
  \bibinfo {pages} {050502} (\bibinfo {year} {2010})}\BibitemShut {NoStop}%
\bibitem [{\citenamefont {Asb\'oth}\ and\ \citenamefont
  {Obuse}(2013)}]{PhysRevB.88.121406}%
  \BibitemOpen
  \bibfield  {author} {\bibinfo {author} {\bibfnamefont {J.~K.}\ \bibnamefont
  {Asb\'oth}}\ and\ \bibinfo {author} {\bibfnamefont {H.}~\bibnamefont
  {Obuse}},\ }\href {https://doi.org/10.1103/PhysRevB.88.121406} {\bibfield
  {journal} {\bibinfo  {journal} {Phys. Rev. B}\ }\textbf {\bibinfo {volume}
  {88}},\ \bibinfo {pages} {121406(R)} (\bibinfo {year} {2013})}\BibitemShut
  {NoStop}%
\bibitem [{\citenamefont {Yamagishi}\ \emph
  {et~al.}(2023{\natexlab{b}})\citenamefont {Yamagishi}, \citenamefont
  {Hatano}, \citenamefont {Imura},\ and\ \citenamefont
  {Obuse}}]{PhysRevA.107.042206}%
  \BibitemOpen
  \bibfield  {author} {\bibinfo {author} {\bibfnamefont {M.}~\bibnamefont
  {Yamagishi}}, \bibinfo {author} {\bibfnamefont {N.}~\bibnamefont {Hatano}},
  \bibinfo {author} {\bibfnamefont {K.-I.}\ \bibnamefont {Imura}},\ and\
  \bibinfo {author} {\bibfnamefont {H.}~\bibnamefont {Obuse}},\ }\href
  {https://doi.org/10.1103/PhysRevA.107.042206} {\bibfield  {journal} {\bibinfo
   {journal} {Phys. Rev. A}\ }\textbf {\bibinfo {volume} {107}},\ \bibinfo
  {pages} {042206} (\bibinfo {year} {2023}{\natexlab{b}})}\BibitemShut
  {NoStop}%
\bibitem [{\citenamefont {Yao}\ and\ \citenamefont
  {Wang}(2018)}]{PhysRevLett.121.086803}%
  \BibitemOpen
  \bibfield  {author} {\bibinfo {author} {\bibfnamefont {S.}~\bibnamefont
  {Yao}}\ and\ \bibinfo {author} {\bibfnamefont {Z.}~\bibnamefont {Wang}},\
  }\href {https://doi.org/10.1103/PhysRevLett.121.086803} {\bibfield  {journal}
  {\bibinfo  {journal} {Phys. Rev. Lett.}\ }\textbf {\bibinfo {volume} {121}},\
  \bibinfo {pages} {086803} (\bibinfo {year} {2018})}\BibitemShut {NoStop}%
\bibitem [{\citenamefont {Yokomizo}\ and\ \citenamefont
  {Murakami}(2019)}]{PhysRevLett.123.066404}%
  \BibitemOpen
  \bibfield  {author} {\bibinfo {author} {\bibfnamefont {K.}~\bibnamefont
  {Yokomizo}}\ and\ \bibinfo {author} {\bibfnamefont {S.}~\bibnamefont
  {Murakami}},\ }\href {https://doi.org/10.1103/PhysRevLett.123.066404}
  {\bibfield  {journal} {\bibinfo  {journal} {Phys. Rev. Lett.}\ }\textbf
  {\bibinfo {volume} {123}},\ \bibinfo {pages} {066404} (\bibinfo {year}
  {2019})}\BibitemShut {NoStop}%
\bibitem [{\citenamefont {Kawabata}\ \emph {et~al.}(2020)\citenamefont
  {Kawabata}, \citenamefont {Okuma},\ and\ \citenamefont
  {Sato}}]{PhysRevB.101.195147}%
  \BibitemOpen
  \bibfield  {author} {\bibinfo {author} {\bibfnamefont {K.}~\bibnamefont
  {Kawabata}}, \bibinfo {author} {\bibfnamefont {N.}~\bibnamefont {Okuma}},\
  and\ \bibinfo {author} {\bibfnamefont {M.}~\bibnamefont {Sato}},\ }\href
  {https://doi.org/10.1103/PhysRevB.101.195147} {\bibfield  {journal} {\bibinfo
   {journal} {Phys. Rev. B}\ }\textbf {\bibinfo {volume} {101}},\ \bibinfo
  {pages} {195147} (\bibinfo {year} {2020})}\BibitemShut {NoStop}%
\end{thebibliography}%

\end{document}